\documentclass[aps,twocolumn,10pt,floatfix,secnumarabic,superscriptaddress]{revtex4}

\usepackage{graphicx}
\usepackage{amsmath}
\usepackage{amssymb}
\usepackage[sort&compress]{natbib}
\usepackage{mathrsfs}

\DeclareRobustCommand{\<}[1]{\hspace{-0.11111em}#1\hspace{-0.11111em}}
\DeclareRobustCommand{\rtrim}[1]{#1\hspace{-0.11111em}}
\DeclareRobustCommand{\abs}[1]{\lvert#1\rvert}
\usepackage{closedcases}
\usepackage{liealg}
\DeclareRobustCommand{\grpsochain}{\grpso{5}\<\supset\grpso{3}}
\DeclareRobustCommand{\grpsutimes}{\grpsu{1,1}\times\grpso{5}}
\usepackage{stackrel}
\DeclareRobustCommand{\subqn}[2]{\stackrel[#2]{}{#1}}
\DeclareRobustCommand{\subqnphantom}[2]{\stackrel[#2]{}{#1\mathrel{\makebox[0pt]{\phantom{()}}}}}

\DeclareRobustCommand{\Lambdahat}{\hat{\Lambda}}
\DeclareRobustCommand{\bbR}{\mathbb{R}}

\DeclareRobustCommand{\scrN}{\mathscr{N}}
\DeclareRobustCommand{\calQ}{\mathcal{Q}}

\DeclareRobustCommand{\scrD}{\mathscr{D}}

\DeclareRobustCommand{\deltasqrt}[1]{(1+\delta_{#1})^{1/2}}
\DeclareRobustCommand{\deltasqrtinv}[1]{(1+\delta_{#1})^{-1/2}}
\DeclareRobustCommand{\Nmax}{{N_\text{max}}}
\DeclareRobustCommand{\vmax}{{v_\text{\rm max}}}  
\DeclareRobustCommand{\Lmax}{{L_\text{\rm max}}}
\DeclareRobustCommand{\Lmaxnaught}{{L_{\text{\rm max},0}}}

\DeclareRobustCommand{\eig}{{e^{i\gamma}}}
\DeclareRobustCommand{\epig}{{e^{+i\gamma}}}
\DeclareRobustCommand{\emig}{{e^{-i\gamma}}}

\DeclareRobustCommand{\rmh}{\textrm{\-}}
\DeclareRobustCommand{\hideargs}[1]{}

\begin{document}


\title{\boldmath 
Construction of $\grpsochain$ spherical harmonics and Clebsch-Gordan
coefficients }

\author{M. A. Caprio}
\affiliation{Department of Physics, University of Notre Dame,
Notre Dame, Indiana 46556-5670, USA}
\author{D. J. Rowe}
\affiliation{Department of Physics, University of Toronto, Toronto, Ontario M5S\,1A7, Canada}
\author{T. A. Welsh}
\affiliation{Department of Physics, University of Toronto, Toronto, Ontario M5S\,1A7, Canada}

\date{\today}

\maketitle

\onecolumngrid

\noindent\rule{\hsize}{1pt}

~

\noindent\textbf{Abstract}

~

The $\grpsochain$ spherical harmonics form a natural
basis for expansion of nuclear collective model angular wave
functions.  They underlie the recently-proposed algebraic method for
diagonalization of the nuclear collective model Hamiltonian in an
$\grpsutimes$ basis.  We present a computer code for explicit
construction of the $\grpsochain$ spherical harmonics
and use them to compute the Clebsch-Gordan coefficients needed for
collective model calculations in an $\grpso{3}$-coupled basis.  With
these Clebsch-Gordan coefficients it becomes possible to compute the
matrix elements of collective model observables by purely algebraic
methods.

~

\noindent\textbf{Program Summary}

~

\noindent\textit{Title of program:} GammaHarmonic

\noindent\textit{Catalogue identifier:} AECY\_v1\_0

\noindent\textit{Program summary URL:} http://cpc.cs.qub.ac.uk/summaries/AECY\_v1\_0

\noindent\textit{Program available from:} CPC Program Library, Queen's University of Belfast, N. Ireland

\noindent\textit{Operating system:} Any which supports Mathematica;
tested under Microsoft Windows XP and Linux

\noindent\textit{Programming language used:} Mathematica 6

\noindent\textit{Number of bytes in distributed program, including test code and documentation:} 16\,037\,234

\noindent\textit{Distribution format:} tar.gz

\noindent\textit{Nature of problem:} Explicit construction of
$\grpsochain$ spherical harmonics on $S_4$.  Evaluation of
$\grpso{3}$-reduced matrix elements and $\grpsochain$ Clebsch-Gordan
coefficients (isoscalar factors).

\noindent\textit{Method of solution:} Construction of
$\grpso{5}\<\supset\grpso{3}$ spherical harmonics by
orthonormalization, obtained from a generating set of functions,
according to the method of D.~J.~Rowe, P.~S.~Turner, and J.~Repka
[J.~Math.~Phys.~45 (2004) 2761].  Matrix elements and Clebsch-Gordan
coefficients follow by construction and integration of $\grpso{3}$
scalar products.

~


~


~

\noindent\rule{\hsize}{1pt}

~

\twocolumngrid
\newpage  


\section{Introduction}
\label{sec-intro}

The $\grpsochain$ spherical harmonics constitute the natural basis for
the ``angular'' wave functions in the collective model of nuclear
quadrupole motion~\cite{bohr1998:v2}.  The $\grpsochain$ spherical
harmonics underlie the recently-proposed algebraic
scheme~\cite{rowe2004:tractable-collective,rowe2005:algebraic-collective,rowe2005:radial-me-su11}
for the nuclear collective model.  Direct products of the
$\grpsochain$ spherical harmonics with appropriate optimal radial wave
functions provide an $\grpsutimes$ algebraic
basis~\cite{rowe2004:tractable-collective,rowe2005:algebraic-collective,rowe2005:radial-me-su11}
which allows for exceedingly efficient numerical diagonalization of
nuclear collective model Hamiltonians.

For applications to transitional and deformed nuclei, the
$\grpsutimes$ scheme reduces by orders of
magnitude~\cite{rowe2005:algebraic-collective} the basis size needed
for convergence as compared to conventional diagonalization in a
five-dimensional oscillator [$\grpu{5}\<\supset\grpso{5}$]
basis~\cite{gneuss1969:gcm,gneuss1970:gcm,eisenberg1987:v1}.  Matrix
elements of an essentially unlimited set of potential and kinetic
energy operators, often in analytic form, can easily be constructed in
an $\grpsutimes$
basis~\cite{rowe2005:algebraic-collective,rowe2005:radial-me-su11}, in
terms of $\grpsochain$ Clebsch-Gordan coefficients. The resulting calculational scheme, the so-called
algebraic collective model (ACM), is described in
Refs.~\cite{rowe2004:tractable-collective,rowe2005:algebraic-collective,rowe2005:radial-me-su11,rowanwood-INPRESS}.
Examples of physical applications may be found in
Ref.~\cite{caprio2005:axialsep}.

In this article, we present a computer code for explicit construction
of the $\grpsochain$ spherical harmonics and for using them to
determine the Clebsch-Gordan coefficients for coupling of symmetric
irreducible representations of $\grpso{5}$ in an $\grpso{3}$ basis.
The construction of the $\grpsochain$ spherical harmonics is carried
out by orthonormalization of monomials in 
a set of four generating
functions, according to the method of Rowe, Turner, and
Repka~\cite{rowe2004:spherical-harmonics}. The purpose of this code is
to make calculations using the ACM routinely possible, without the
need to reconstruct the $\grpso{5}$ machinery for each new
application.  The $\grpsochain$ Clebsch-Gordan coefficients yielded by
the code are also relevant to the $\grpu{6}$ interacting boson model
(IBM)~\cite{iachello1987:ibm,iachello1991:ibfm}  which, in
$\grpu{6}\<\supset\grpu{5}\<\supset\grpso{5}$ and
$\grpu{6}\<\supset\grpso{6}\<\supset\grpso{5}$ bases, 
is in close correspondence with the collective model in appropriate
$\grpsutimes$ bases~\cite{rowe2005:ibm-geometric}.  They can,
furthermore, be used as seed coefficients in the generation of
Clebsch-Gordan coefficients for the coupling of more general
(non-symmetric) representations of $\grpso{5}$ or $\grpsp{4}$ in an
$\grpso{3}$ basis~\cite{caprio2007:geomsuper2}.

The code accompanying this article is implemented in Mathematica
6~\cite{wolfram2007:mathematica-6}.  All calculations are carried out
in exact symbolic arithmetic.  A machine-readable tabulation of
calculated Clebsch-Gordan coefficients is included along with the code
in the CPC Program Library.  This tabulation is of sufficient extent
to support basic calculations using the algebraic collective model,
without the necessity of re-running the code.

The basic algorithm used here is that presented in
Ref.~\cite{rowe2004:spherical-harmonics}.  However, the computational
techniques have been substantially developed.  For example,
integration via the Fourier representation of functions has been used
to greatly increase the efficiency of the calculations and
considerably extend the practical range of application of the
algorithm.

The necessary mathematical definitions for the
$\grpsochain$ spherical harmonics and the general method for
construction of the basis and evaluation of Clebsch-Gordan
coefficients are discussed in Sec.~\ref{sec-method}.  The more
technical details of the computer implementation are summarized in
Sec.~\ref{sec-impl}.  Instructions for installation and use of the
computer code are given in Sec.~\ref{sec-use}.

\section{The basic algorithm}
\label{sec-method}

\subsection{\boldmath The $\grpsochain$ spherical harmonics}
\label{sec-method-harm}

The $\grpso{5}$ spherical harmonics are eigenfunctions of the
Laplace-Beltrami operator $\Lambdahat^2$ on the four-sphere $S_4$,
that is, the angular part of the Laplacian in five dimensions.  This
operator $\Lambdahat^2$ is also the second order Casimir invariant of
$\grpso{5}$.  Thus, the spherical harmonics are functions of a set of
coordinates on $S_4$.  Standard $(\gamma,\Omega)$ coordinates for
$S_4$ are reviewed below in Sec.~\ref{sec-method-func}.  The
$\grpso{5}$ spherical harmonics constitute a complete orthonormal
basis for the space $L^2(S_4)$ of square-integrable functions on $S_4$
and transform under $\grpso{5}$ rotations as bases for the symmetric
irreps $(v,0)$, for $v\<=0$, $1$, $\ldots$, of $\grpso{5}$.

For applications to the nuclear collective model, we seek spherical
harmonics which have ``good $\grpso{3}$ angular momentum'', that is,
which also transform as bases for irreps, labelled by $(L)$, of the
$\grpso{3}$ subalgebra of $\grpso{5}$.  The desired $\grpsochain$
spherical harmonics, which we denote by $\Psi_{v\alpha{}LM}
(\gamma,\Omega)$, thus reduce the subalgebra chain
\begin{equation}
\label{eqn-sochain}
\subqn{\grpso{5}}{v}
\subqnphantom{\supset}{\alpha}
\subqn{\grpso{3}}{L}
\supset
\subqn{\grpso{2}}{M},
\end{equation}
with the representation labels as shown.  The
label $v$ is conventionally termed the ``boson seniority'' quantum
number, following Racah.  It is
the five-dimensional [$\grpso{5}$] analog of the angular
momentum quantum number.  The $\grpso{5}$ spherical harmonics
satisfy the eigenvalue equation
\begin{equation}
\Lambdahat^2 \Psi_{v\alpha{}LM}
 (\gamma,\Omega) = v (v+3) \Psi_{v\alpha{}LM}
(\gamma,\Omega).
\end{equation}
They are also eigenfunctions of the Casimir operators of the other
algebras in the chain~(\ref{eqn-sochain}).  However, the 
labels $v LM$ provided by these Casimir operators are insufficient to fully
distinguish the spherical harmonics.  In particular, multiple
$\grpso{3}$ representations of the same $L$ may occur within a given
$\grpso{5}$ representation.  The ``missing label'' is provided by a
multiplicity index $\alpha$.  The branching rule for $\grpso{3}$
irreps occurring within an $\grpso{5}$ irrep is
well known~\cite{kemmer1968:so5-irreps-1,williams1968:so5-irreps-2},
and the multiplicity is given by~(\ref{eqn-so5-dvL}).

The $\grpsochain$ spherical harmonics have physical significance as
the angular, \textit{i.e.}, $(\gamma,\Omega)$, wave functions for the
nuclear collective model, for the case in which which the collective
potential is $\grpso{5}$
invariant~\cite{wilets1956:oscillations,bes1959:gamma}.  Recall the
quantum mechanics of a particle in three-dimensional Euclidean space,
subject to a central force, \textit{i.e.}, in spherical coordinates,
$V(r,\theta,\varphi)\<\rightarrow V(r)$.  The Hamiltonian is then
$\grpso{3}$ invariant, and its eigenfunctions, with good angular
momentum quantum numbers, factorize into products $f_{nl}(r) Y_{lm}
(\theta,\varphi)$ of radial wave functions and $\grpso{3}$ spherical
harmonics.  Similarly, for the Bohr Hamiltonian, which
is given in terms of quadrupole deformation variables $\beta$,
$\gamma$, and $\Omega$ (Sec.~\ref{sec-method-func}) by
\begin{equation}
-\frac{\hbar^2}{2B}\Bigl[
\frac{1}{\beta^4}\frac{\partial}{\partial\beta}\beta^4\frac{\partial}{\partial\beta}
-\frac{\Lambdahat^2}{\beta^2}\Bigr]+V(\beta,\gamma),
\end{equation}
if the potential is a function $V(\beta)$ of the radial coordinate
only, then the five-dimensional analog of a central force problem
arises.  The Hamiltonian is $\grpso{5}$ invariant, and its
eigenfunctions, with good seniority and angular momentum quantum
numbers, factorize into products $f_{nv}(\beta)
\Psi_{v\alpha{}LM} (\gamma,\Omega)$ of radial, \textit{i.e.}, $\beta$, wave functions and
$\grpso{5}$ 
spherical harmonics, as in Ref.~\cite{wilets1956:oscillations}.
An example of application of the present methods to such ``$\gamma$-unstable''
problems is found in Ref.~\cite{caprio2007:geomsuper2}.

A limited set of $\grpsochain$ spherical harmonics was computed many
years ago by B\`es~\cite{bes1959:gamma}, for values of the angular
momentum $L\leq6$.  These were obtained by series solution of coupled
differential equations in the coordinate $\gamma$, to find the
eigenfunctions of the $\grpso{5}$ Casimir invariant.  However, this
approach becomes prohibitively complicated for $L>6$.  The
$\grpsochain$ spherical harmonics are the angular wave functions for
the five-dimensional quadrupole harmonic
oscillator~\cite{chacon1976:oscillator,chacon1977:oscillator}.
Thus, alternative approaches based on the oscillator basis construction are
possible, as summarized in Ref.~\cite{eisenberg1987:v1}.  A method
based on the Cartan-Weyl reduction is given in
Ref.~\cite{debaerdemacker2007:so5-cartan-weyl}.

Here we make use of the construction proposed in
Ref.~\cite{rowe2004:spherical-harmonics}, in which the $\grpsochain$
spherical harmonics are developed as polynomials in a set of four
basic generating functions defined on $S_4$.  These functions were
identified as generators of a complete linearly independent basis of
$\grpso{3}$-coupled wave functions for $L^2(S_4)$~\cite{fn-integrity},
based on a knowledge of $L^2(S_4)$ as a direct sum of irreducible
$\grpso{3}$ subspaces [obtained from the
$\grpso{5}\<\rightarrow\grpso{3}$ branching rules].  The generating
functions can be related to the ``elementary permissible diagrams" of
the well-known algorithm of Chac\'on
\textit{et al.}~\cite{chacon1976:oscillator,chacon1977:oscillator} for
construction of a $\grpu{5}\<\supset\grpso{5}\<\supset\grpso{3}$ basis
for the five-dimensional harmonic oscillator, implemented in the
nuclear collective model code of Hess \textit{et
al.}~\cite{hess1980:gcm-details-238u,troltenier1991:gcm}.  The salient
property of the generating
functions for $L^2(S_4)$~\cite{rowe2004:spherical-harmonics} is that
they yield a direct route to the construction of $\grpsochain$
spherical harmonics without reference to
the full harmonic oscillator problem.
The method is set forth more concretely in
Sec.~\ref{sec-method-basis}.

\subsection{\boldmath Representation of functions on $S_4$}
\label{sec-method-func}

The quadrupole moments $q_m$ ($m\<=0$, $\pm1$,
$\pm2$) for the collective model transform as 
components of an $L=2$ spherical tensor under $\grpso{3}$ rotations, 
\textit{i.e.},
\begin{equation}
q_m \rightarrow \sum_k q_k\scrD^{(2)}_{km} (\Omega) ,
\end{equation}
where $\Omega$ represents the Euler angles for the $\grpso{3}$
rotation, and the Wigner $\scrD$ function~\cite{edmonds1960:am} is the
rotation matrix element.  [The
quantities $q_m$ can alternatively be taken to represent the
nuclear surface deformation parameters
$\alpha_m$ ($m\<=0$, $\pm1$, $\pm2$).  The difference is only in
physical interpretation and does not affect the following results for
$\grpso{5}$.]  

These quadrupole moments
 are conveniently expressed in terms of Bohr's spherical polar
coordinates $(\beta,\gamma,\Omega)$~\cite{bohr1952:vibcoupling}, by
the relation
\begin{equation}
q_m = \beta \cos\gamma\, \scrD^{(2)}_{0,m}(\Omega) +
\frac{1}{\sqrt{2}} \beta\sin\gamma \Bigl[ \scrD^{(2)}_{2,m}(\Omega) +
\scrD^{(2)}_{-2,m}(\Omega) \Bigr]. 
\end{equation}
The squared length of a vector $q \in \bbR^5$ is given by $\sum_m
\abs{q_m}^2\<=\beta^2$.  Thus, $\beta$ is the radial coordinate for
$\bbR^5$, and $(\gamma,\Omega)$ are angular coordinates.

For consideration of the angular functions 
on the unit sphere $S_4$, we henceforth  set $\beta\<=1$ and restrict consideration to
the unit length quadrupole moments $\calQ_m$, defined as
\begin{equation}
\calQ_m =   \cos\gamma\, \scrD^{(2)}_{0,m}(\Omega) 
+
\frac{1}{\sqrt{2}} \sin\gamma \Bigl[ \scrD^{(2)}_{2,m}(\Omega) +
\scrD^{(2)}_{-2,m}(\Omega) \Bigr].
\end{equation}
These unit-length quadrupole moments are then proportional to the
basic $v\<=1$ spherical harmonics on $S_4$.

Consider a function $\Psi^{(L)}_M(\gamma,\Omega)$ on $S_4$, of good
$\grpso{3}$ angular momentum $L$ and $\grpso{2}$ quantum number $M$.
Any such function may be expanded~\cite{bohr1952:vibcoupling} in the form
\begin{equation}
\label{eqn-expansion-m}
\Psi^{(L)}_M(\gamma,\Omega)=\sum_{\substack{K=0\\\text{even}}}^L
F_K(\gamma) \xi^{(L)}_{KM}(\Omega),
\end{equation}
where
\begin{equation} 
\label{eqn-xi-m}
\xi^{(L)}_{KM}(\Omega) \equiv \frac{1}{\deltasqrt{K}}
\Bigl[\scrD^{(L)}_{KM}(\Omega)+(-)^L\scrD^{(L)}_{-KM}(\Omega)\Bigr].
\end{equation} 
Note that the $\scrD$ functions occur only in the 
symmetrized linear combinations
$\scrD^{(L)}_{KM}+(-)^L\scrD^{(L)}_{-KM}$
with even values of $K$.  Thus,
 because $\xi^{(L)}_{-KM} = (-)^L \xi^{(L)}_{KM}$,
we can restrict to $K\geq 0$.
Note also that the functions $\xi^{(L)}_{KM}$ vanish identically for $K=0$ when $L$ is odd.
The normalization factor $\deltasqrtinv{K}$ (where
$\delta_K\<\equiv\delta_{K,0}$) is included for later convenience.

The functions $\xi^{(L)}_{KM}(\Omega)$,  with $K\geq 0$, provide an orthogonal basis for those functions
of the $\grpso{3}$ angles which respect the symmetry
properties of collective model wave functions.   From
the inner product for the
$\scrD$ functions~\cite{edmonds1960:am},
\begin{equation}
\int \scrD^{(L')\,*}_{K'M'}  (\Omega) \scrD^{(L)}_{KM}  (\Omega) \,d\Omega
\<= \frac{8\pi^2}{2L+1}\delta_{L'L} \delta_{K'K} \delta_{M'M},
\end{equation}
 we obtain the inner product for the nonzero $\xi^{(L)}_{KM} (\Omega)$
\begin{equation}
\int \xi^{(L')\,*}_{K'M'}  (\Omega)\xi^{(L)}_{KM} (\Omega) \, d\Omega = \frac{16 \pi^2}{2L+1}
\delta_{L'L} \delta_{K'K} \delta_{M'M}.  
\end{equation}
The volume element on
$S_4$ is given by $dv\<=\sin 3\gamma\,d\gamma\,d\Omega$, where
$\gamma$ is integrated over the range $[0,\pi/3]$.  Thus, the inner product $\ol{\Psi_2}{\Psi_1}\<\equiv\int
\Psi_2^*\Psi_1 \sin3\gamma\,d\gamma\,d\Omega$ of two  functions on $S_4$, when expanded
according to~(\ref{eqn-expansion-m}), is given simply by 
\begin{equation}
\label{eqn-overlap-sum}
\ol{\Psi^{(L)}_{2\,M}}{\Psi^{(L)}_{1\,M}} 
=
\frac{16 \pi^2}{2L+1} 
\!
\int
\!
\Biggl[ 
\sum_{\substack{K=0\\\text{even}}}^L
F_{1\,K}(\gamma)F_{2\,K}(\gamma)
\Biggr]
\hspace{-0.3ex}
\sin 3\gamma\,d\gamma
.
\end{equation}
Functions of distinct $L$ or $M$ are orthogonal.

\subsection{\boldmath$\grpso{3}$-coupled  basis functions} 
\label{sec-method-basis}

The 
generating function method~\cite{rowe2004:spherical-harmonics} for
construction of the $\grpsochain$ spherical harmonics rests upon the
observation that a complete set of normalizable $\grpso{3}$ highest-weight
(\textit{i.e.}, those for which $M\<=L$) functions on $S_4$ is provided by the products
\begin{multline}
\label{eqn-phi-prod}
\Phi_{[n_1,n_2,n_3,n_4]}(\gamma,\Omega)\\
= [\Phi_1(\gamma,\Omega)]^{n_1} [\Phi_2(\gamma,\Omega)]^{n_2}
[\Phi_3(\gamma,\Omega)]^{n_3} [\Phi_4(\gamma,\Omega)]^{n_4},
\end{multline}
where the exponents $n_1$, $n_2$, and $n_3$ take on the values $0$,
$1$, $\ldots$, and $n_4$ is restricted to $0$ or $1$.  These
$\Phi_{[n_1,n_2,n_3,n_4]}$ are monomials in four generating functions
$\Phi_1$, $\Phi_2$, $\Phi_3$, and $\Phi_4$,
which are simply the $M=L$ components 
\begin{equation}
\label{eqn-phi-fcn}
\begin{aligned}
\Phi_1&\propto\calQ_2 \\
\Phi_2&\propto(\calQ\times\calQ)^{(2)}_2\\
\Phi_3&\propto(\calQ\times\calQ\times\calQ)^{(0)}_0\\
\Phi_4&\propto(\calQ\times\calQ\times\calQ)^{(3)}_3,
\end{aligned}
\end{equation}
of $\grpso{3}$-coupled products involving $\calQ$.
The coupled product of two tensors is defined by
\begin{equation}
\label{eqn-tensor-couple}
[U^{(L_2)}\times T^{(L_1)}]^{(L)}_M
\equiv\sum_{M_1M_2} \smallCG{L_1}{M_1}{L_2}{M_2}{L}{M} 
U^{(L_2)}_{M_2} T^{(L_1)}_{M_1} .
\end{equation}
Note the
right-to-left coupling order in this definition of the $\grpso{3}$
tensor product, used for consistency with
Refs.~\cite{rowe2004:spherical-harmonics,rowanwood-INPRESS}  and
to simplify the phases~\cite{fn-coupling} arising in the
Wigner-Eckart theorem.
The norms of these generating functions are
arbitrary  and can be chosen for convenience.  
In the standard form~(\ref{eqn-expansion-m}), we take 
\begin{equation}
\label{eqn-phi-expansion}
\begin{alignedat}{4}
\Phi_1(\gamma,\Omega)&=\cos\gamma &\,&\xi^{(2)}_{02}(\Omega) &\,+\,& \sin\gamma &\,&\xi^{(2)}_{22}(\Omega)\\
\Phi_2(\gamma,\Omega)&=\cos2\gamma &\,&\xi^{(2)}_{02}(\Omega) &\,-\,& \sin2\gamma &\,&\xi^{(2)}_{22}(\Omega)\\
\Phi_3(\gamma,\Omega)&=\cos3\gamma &\,&\xi^{(0)}_{00}(\Omega)&&\\
\Phi_4(\gamma,\Omega)&= &&&&\sin3\gamma &\,&\xi^{(3)}_{23}(\Omega).
\end{alignedat}
\end{equation}

The multiplication of two highest-weight functions yields another
highest-weight function (equivalent to the stretched coupling of two angular
momentum functions).  Hence, 
 every $\Phi_{[n_1,n_2,n_3,n_4]}$ is a
highest-weight function, with 
$L\<=M\<=2n_1+2n_2+3n_4$
and ``degree" $N\<=n_1+2n_2+3n_3+3n_4$ in
the unit quadrupole moments $\calQ$~\cite{fn-degree}.
The
multiplication of the basis functions is carried out using the
multiplication rule for the highest-weight functions
$\xi^{(L)}_{KL}$. 
This rule, which follows from the multiplication rule~\cite{edmonds1960:am} for 
$\scrD$ functions 
\begin{multline}
\label{eqn-cg-series}
\scrD^{(L_2)}_{K_2M_2}(\Omega) \scrD^{(L_1)}_{K_1M_1}(\Omega) \\
= \sum_{L} \smallCG{L_1}{K_1}{L_2}{K_2}{L}{K}
\smallCG{L_1}{M_1}{L_2}{M_2}{L}{M}
\scrD^{(L)}_{KM}(\Omega) , 
\end{multline}
where $K=K_1+K_2$ and $M=M_1+M_2$,
is expressed by the equation  
\begin{multline}
\label{eqn-xi-prod-stretched}
\xi^{(L_2)}_{K_2L_2} \xi^{(L_1)}_{K_1L_1} 
= \sum_{K\geq 0} \frac{\deltasqrt{K}}{\deltasqrt{K_1}\deltasqrt{K_2}}
\\\times
\Biggl[
\smallCG{L_1}{K_1}{L_2}{K_2}{L_1+L_2}{K}
+ (-)^{L_2} \smallCG{L_1}{K_1}{L_2}{-K_2}{L_1+L_2}{K}
\Biggr] \xi^{L_1+L_2}_{K,L_1+L_2} .
\end{multline}
The more general tensor-coupled product of the spherical tensors
$\xi^{(L)}_{K}$, which have components $\xi^{(L)}_{KM}$, is considered
in Sec.~\ref{sec-method-me}.
\begin{figure}
\begin{center}
\includegraphics*[width=\hsize]{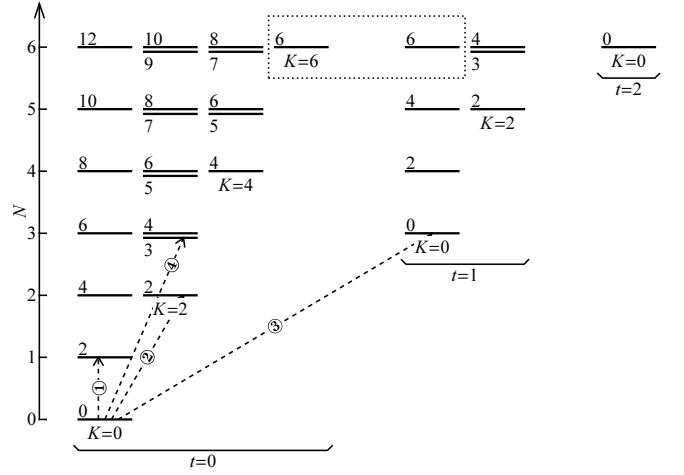}
\end{center}
\caption{The set of basis monomials
$\Phi_{NtL}$ of degree $N\<\leq6$, with labels $N$, $L$, $t$, and $K$
indicated.  The numbered arrows indicate the raising actions of
multiplication by the generating functions $\Phi_1$, $\Phi_2$,
$\Phi_3$, and $\Phi_4$.  The angular momentum multiplicity at $N\<=6$
is highlighted (dotted box).  This same diagram enumerates the
$\Psi_{v\alpha L}$ with $v\<\leq6$, that is, it gives the branching of
$\grpso{5}$ irreps $(v,0)$ into angular momenta $L$, if the axis label
$N$ is read as $v$.}
\label{fig-phi}
\end{figure}
\begin{table*}
\caption{The $\grpsochain$ spherical harmonics $\Psi_{v\alpha L}$ for
$v\<\leq3$.  For these lowest-seniority spherical harmonics,
orthonormalization is trivial, and the spherical harmonics are simply
proportional to the corresponding monomial basis members
($\Psi_{v\alpha L}=\scrN_{v\alpha L}^{-1/2}\Phi_{NtL}$).  The relations
between the $\Psi_{v\alpha L}$ (with $L\<=0$ and $2$) and
commonly-encountered scalar and quadrupole operators are given in
brackets.}
\label{tab-psi}
\begin{ruledtabular}
\renewcommand{\arraystretch}{2}
\begin{tabular}{rrrl}
$v$ & $L$ & $\scrN_{v\alpha{}L}/(8\pi^2)$ & $(8\pi^2)^{1/2}\Psi_{v\alpha{}L}$\\
\hline
0 & 0 & 2/3 & $\tfrac{\sqrt{3}}{2} \,\xi^{(0)}_0$ \quad $\left[=\sqrt{\tfrac{3}{2}}\right]$\\
1 & 2 &  4/15 & $\tfrac{\sqrt{15}}{2} \cos\gamma\,\xi^{(2)}_0 + \tfrac{\sqrt{15}}{2} \sin\gamma\,\xi^{(2)}_2 $\quad $\left[=\sqrt{\tfrac{15}{2}}\calQ\right]$\\
2 & 4 &  16/105 & $\tfrac{\sqrt{3}}{8}(7+5\cos2\gamma) \,\xi^{(4)}_0 +\tfrac{3\sqrt{5}}{4} \sin2\gamma\,\xi^{(4)}_2 +\tfrac{\sqrt{105}}{8} (1-\cos2\gamma)\,\xi^{(4)}_4$          \\
  & 2 &  4/15 & $\tfrac{\sqrt{15}}{2} \cos2\gamma\,\xi^{(2)}_0 - \tfrac{\sqrt{15}}{2} \sin2\gamma\,\xi^{(2)}_2 $\quad $\left[=-\tfrac{\sqrt{105}}{2}(\calQ\times\calQ)^{(2)}\right]$           \\
3 & 6 &  32/315 & $\tfrac{9}{16}\sqrt{\tfrac{5}{11}}(7\cos\gamma+\cos3\gamma)\,\xi^{(6)}_0 + \tfrac{3}{16}\sqrt{\tfrac{7}{22}}(15\sin\gamma+11\sin3\gamma)\,\xi^{(6)}_2  $ \\
&&&\quad $+\tfrac{9}{16}\sqrt{\tfrac{35}{11}}(\cos\gamma-\cos3\gamma)\,\xi^{(6)}_4 + \tfrac{3}{16}\sqrt{\tfrac{35}{2}}(\sin\gamma-\sin3\gamma)\,\xi^{(6)}_6$             \\
  & 4 &  88/945 & $\tfrac{3}{4}\sqrt{\tfrac{3}{22}}(5\cos\gamma+7\cos3\gamma) \,\xi^{(4)}_0 -\tfrac{9}{2}\sqrt{\tfrac{5}{22}} \sin\gamma\,\xi^{(4)}_2 +\tfrac{3}{4}\sqrt{\tfrac{105}{22}}(-\cos\gamma+\cos3\gamma)\,\xi^{(4)}_4$                        \\
  & 3 &  8/63 & $\tfrac{3}{2}\sqrt{\tfrac{7}{2}}\sin3\gamma\,\xi^{(3)}_2$            \\
  & 0 &  4/9 &  $\tfrac{3}{2}\cos3\gamma\,\xi^{(0)}_0$\quad $\left[=\tfrac{3}{\sqrt{2}} \cos3\gamma \right]$
\end{tabular}
\end{ruledtabular}
\end{table*}


An equivalent but more descriptive labeling for the highest-weight
monomials $\Phi_{[n_1,n_2,n_3,n_4]}(\gamma,\Omega)$ is given by
regarding each of them as the $M\<=L$ component of the set of angular
momentum $L$ functions $\Phi_{NtLM}(\gamma,\Omega)$ to which they
naturally extend.  
Here monomials of the same degree $N$
and angular momentum $L$ are distinguished by the label $t\<\equiv
n_3$, which counts the number of zero-coupled triplets of the
quadrupole coordinates.
In addition, the $\Phi$ may be organized into quasibands of given
bandhead angular momentum $K\<=2n_2+2n_4$, as shown in
Fig.~\ref{fig-phi}.
Together these functions $\Phi_{NtLM}(\gamma,\Omega)$ constitute a spherical
tensor $\Phi_{NtL}(\gamma,\Omega)$.  Once the highest weight function
$\Phi_{NtLL}(\gamma,\Omega)$ has been defined, the remaining
$\Phi_{NtLM}(\gamma,\Omega)$
follow immediately, since they share the same coefficients
$F_K(\gamma)$ in the expansion~(\ref{eqn-expansion-m}).

An important characteristic of the generating function construction
is that the set of $L$ values appearing at a given degree $N$ is
identical to the set of $L$ values arising for $\grpso{3}$ irreps in
an $\grpso{5}$ irrep of seniority $v=N$. These are given by the known
$\grpsochain$ branching
rule~\cite{kemmer1968:so5-irreps-1,williams1968:so5-irreps-2}.  Thus,
Fig.~\ref{fig-phi} also enumerates the labels of all highest-weight spherical
harmonics $\Psi_{v\alpha LL}$ up to the maximum seniority ($v\<=6$)
shown.  The multiplicity $d_{NL}$ of angular momentum $L$ at degree
$N$ (also the multiplicity $d_{vL}$ of angular momentum $L$ at
seniority $v$) is given by~(\ref{eqn-so5-dvL}).  A multiple occurrence
of the same $L$ at a given $N$ first arises at $N\<=6$, for $L\<=6$
(Fig.~\ref{fig-phi}).

Although the $\Phi_{NtLL}$ form a complete  set of
highest-weight functions and are angular momentum eigenfunctions,
they are, in general, non-orthogonal (for a given $L$) and also not
eigenfunctions of $\Lambdahat^2$.  Thus, they are not the desired
$\grpso{5}$ spherical harmonics.  Construction of the spherical
harmonics $\Psi_{v\alpha LM}(\gamma,\Omega)$ relies on the observation
that the highest-weight spherical harmonics $\Psi_{v\alpha L L}$, like
the $\Phi_{NtLL}$, form a complete set of highest-weight functions and
that they are polynomials of degree $v$ in the unit quadrupole moments
$\calQ$.  For any given $\Nmax$, the sets $\lbrace\Psi_{v\alpha LL}
\vert v\<\leq\Nmax\rbrace$ and $\lbrace\Phi_{NtLL} \vert
N\<\leq\Nmax\rbrace$ both span the space of highest-weight polynomials
of degree $\rtrim\leq\Nmax$ in $\calQ$.  Construction therefore
proceeds inductively, by orthonormalization of the bases of
successively higher $\Nmax$.  The $\Psi_{v\alpha LL}$, as polynomials
of degree $v$ in $\calQ$, are linear combinations of
the $\Phi_{NtLL}$ with $N\<\leq v$.   Because they must be orthogonal to all
$\Psi_{v'\alpha'LL}$ of lower seniority $v'\<<v$,  they can therefore
be obtained by Gram-Schmidt orthogonalizing the monomials of 
degree $N\<=v$ with respect to the space of lower degree.

The orthogonality of the spaces with differing $L$ implies that
orthogonalization can be performed separately in each space of
given $L$.  Within an $L$-space, the spherical
harmonics up to seniority $\vmax$ are obtained as follows:

(1) Order the basis monomials $\Phi_{NtLL}$, for $0\<\leq
N\<\leq\vmax$,  by increasing $N$,  and then by increasing $t$ when multiple basis
monomials of the same $L$ occur for a given $N$~\cite{fn-gst-order-foreshadow}.  These functions may
then be labeled with a single counting index, as $\Phi_{Li}$, with
$i\<=1$, $\ldots$, $D_{\vmax L}$.  The dimension $D_{\vmax L}$ of the
seniority-truncated $L$ space is given by~(\ref{eqn-so5-DvL}).

(2) Calculate the overlaps $\ol{\Phi_{Lj}}{\Phi_{Li}}$ for $1\<\leq i,j\<\leq D_{\vmax L}$,  
using~(\ref{eqn-overlap-sum}). 

(3) Determine the linear  transformation 
necessary to bring the $\Phi_{Li}$ into an orthonormal set by the
Gram-Schmidt procedure.  The result is a matrix of orthogonalization
(and normalization) coefficients $T_{Lij}$ for the $L$-space,
in terms of which the 
\begin{equation}
\label{eqn-gst}
\Psi_{Li}=\sum_{j=1}^{D_{\vmax L}}T_{Lij}\Phi_{Lj}
\end{equation}
are the desired spherical harmonics, in order of increasing seniority.

In step (1), if angular momentum multiplicity occurs at a given $N$,
the ordering of the $\Phi_{NtL}$ sharing the \textit{same} $N$ is, in principle,
arbitrary.  However, choosing a different ordering for the
$\Phi_{NtL}$ gives rise, after Gram-Schmidt orthogonalization, to a
different, equally valid, set of spherical harmonics at the
corresponding seniority ($v\<=N$).  That is, the spherical harmonics
$\Psi_{v\alpha L}$ ($\alpha=1$, $\ldots$, $d_{vL}$) span a
seniority-degenerate subspace, and the ordering of the $\Phi_{NtL}$ in
the orthonormalization process determines which of the possible
unitarily-equivalent bases for the subspace is selected as the
``spherical harmonics''.

Note that the overlaps of the $\Phi_{NtLL}$ obey a parity selection
rule.  Since $\Phi_{NtLL}$ is a product of $N$ factors of $\calQ$, it
has parity $(-)^N$ under the $\bbR^5$ parity operation, which takes
$\calQ_m\<\rightarrow-\calQ_m$.  The overlap of two functions of
opposite $\bbR^5$-parity~\cite{fn-parity}\nocite{pietralla1998:ibm2-d-parity} vanishes, so
$\ol{\Phi_{N't'LL}}{\Phi_{NtLL}}$ is nonzero only if $N+N'$ is even.
The $\Psi_{v\alpha L L}$ resulting from the orthonormalization process
therefore retain definite $\bbR^5$-parity $(-)^v$ and are constructed
only from the $\Phi_{NtLL}$ of this same parity.

Hence, by the $\bbR^5$ parity selection rule, all the $\Phi_{NtLM}$
with $N\<\leq3$ are already orthogonal and therefore \textit{are} the
spherical harmonics with $v\<\leq3$, to within normalization.  The
necessary normalization factors $\scrN_{v\alpha L}$, such that
$\Psi_{v\alpha L}=\scrN_{v\alpha L}^{-1/2}\Phi_{NtL}$, and resulting spherical
harmonics, for $v\<\leq3$, are summarized in Table~\ref{tab-psi}.  The
table also provides a glossary relating the $\Psi_{v\alpha L}$ to Hamiltonian
($L\<=0$) and electric quadrupole ($L\<=2$) operators commonly
referenced in physical applications.  More generally, even for
$v\<>3$, the two lowest-seniority $\Psi_{v\alpha L}$ for a given value
of $L$ are proportional to the $\Phi_{NtL}$ with $N\<=v$.
Other higher-seniority $\Psi_{v\alpha L}$ are nontrivial linear
combinations~(\ref{eqn-gst}) of the $\Phi_{NtL}$ with $N\<\leq v$.

\subsection{Triple overlap integrals} 
\label{sec-method-me}

For the determination of $\grpso{5}$ Clebsch-Gordan coefficients
(Sec.~\ref{sec-cg}), we need to calculate triple overlap integrals
\begin{equation}
\me{\Psi_3}{\hat\Psi_2}{\Psi_1}
\equiv \int 
\Psi^*_3(\gamma,\Omega)\Psi_2(\gamma,\Omega)\Psi_1(\gamma,\Omega) \, \sin
3\gamma\,d\gamma\,d\Omega.
\end{equation}
In this expression, $\hat \Psi_2$ is interpreted as an
operator which acts multiplicatively, \textit{i.e.}, $\hat\Psi_2
\ket{\Psi_1}$ has as its wave function $\Psi_2(\gamma,\Omega)
\Psi_1(\gamma,\Omega)$.  For angular momentum
coupled functions, we need only the reduced matrix elements
$\rme{\Psi^{(L_3)}_{3}}{\hat\Psi^{(L_2)}_{2}}{\Psi^{(L_1)}_{1}}$
defined by the Wigner-Eckart theorem~\cite{edmonds1960:am}
\begin{multline}
\label{eqn-we-defn}
\me{\Psi^{(L_3)}_{3\,M_3}}{\hat\Psi^{(L_2)}_{2\,M_2}}{\Psi^{(L_1)}_{1\,M_1}} = 
\frac{1}{(2L_3+1)^{1/2}}  \smallCG{L_1}{M_1}{L_2}{M_2}{L_3}{M_3}
\\
\times \rme{\Psi^{(L_3)}_{3}}{\hat\Psi^{(L_2)}_{2}}{\Psi^{(L_1)}_{1}} ,
\end{multline}
where the quantity in parentheses is an $\grpso{3}$ Clebsch-Gordan
coefficient.  

The Wigner-Eckart theorem is easily inverted (\textit{e.g.},
Ref.~\cite{rowanwood-INPRESS}) by application of 
the Clebsch-Gordan unitarity condition, to give the reduced matrix
element in a computationally convenient form in terms of the coupled action
of $\hat{\Psi}^{(L_2)}_2$ on $\ket{\Psi^{(L_1)}_1}$.  Namely,
\begin{multline}
\label{eqn-we-overlap}
\rme{\Psi^{(L_3)}_{3}}{\hat\Psi^{(L_2)}_{2}}{\Psi^{(L_1)}_{1}}  
\\
=(2L_3+1)^{1/2}
\bra{\Psi^{(L_3)}_{3\,M_3}} \Big[\hat \Psi^{(L_2)}_2 \times \ket{\Psi^{(L_1)}_1}\Big]^{(L_3)}_{ M_3} ,
\end{multline}
where $\big[\hat\Psi^{(L_2)}_2 \times
\ket{\Psi^{(L_1)}_1}\big]^{(L_3)}_{ M_3}$  has wave function
\begin{multline}
\label{eqn-coupled-wf}
 \Big[\Psi^{(L_2)}_2 (\gamma,\Omega)\times \Psi^{(L_1)}_1 (\gamma,\Omega)\Big]^{(L_3)}_{M_3} 
 = 
\\
\sum_{M_1M_2} \smallCG{L_1}{M_1}{L_2}{M_2}{L_3}{M_3} 
\Psi^{(L_2)}_{2\,M_2} (\gamma,\Omega)\Psi^{(L_1)}_{1\,M_1}(\gamma,\Omega), 
\end{multline}
following the convention~(\ref{eqn-tensor-couple}) for the
coupled product of tensors.
The functions $\xi^{(L)}_{KM}$ ($M=0$, $\pm 1$, $\ldots$, $\pm L$), regarded as components of a spherical tensor $\xi^{(L)}_{K}$, obey the coupling rule
\begin{widetext}
\begin{multline}
\label{eqn-xi-prod}
\bigl[ \xi^{(L_2)}_{K_2}(\Omega) \times \xi^{(L_1)}_{K_1}(\Omega) \bigr]^{(L)}
=
\frac{\deltasqrt{K_1+K_2}}{\deltasqrt{K_1}\deltasqrt{K_2}}
\smallCG{L_1}{K_1}{L_2}{K_2}{L}{K_1+K_2}
\xi^{(L)}_{K_1+K_2}(\Omega)
\\
+
\frac{\deltasqrt{K_1-K_2}}{\deltasqrt{K_1}\deltasqrt{K_2}}
 \begin{closedcases} (-)^{L_2}
 \smallCG{L_1}{K_1}{L_2}{-K_2}{L}{K_1-K_2} \xi^{(L)}_{K_1-K_2}(\Omega) &
 K_1\geq K_2\\ (-)^{L_1} 
 \smallCG{L_1}{-K_1}{L_2}{K_2}{L}{-K_1+K_2}
 \xi^{(L)}_{-K_1+K_2}(\Omega) & K_1\leq K_2
\end{closedcases},
\end{multline}
obtained by direct application of~(\ref{eqn-cg-series}).
It follows that~\cite{fn-rme-sum}
\begin{multline}
\label{eqn-rme-sum}
\rme{\Psi^{(L_3)}_{3}}{\hat \Psi^{(L_2)}_{2}}{\Psi^{(L_1)}_{1}} = 
\frac{16\pi^2}{(2L_3+1)^{1/2}}
\int 
\Biggl[
\sum_{\substack{{K_1,K_2,K_3}\\\text{even}}}\frac{\deltasqrt{K_3}}{\deltasqrt{K_1}\deltasqrt{K_2}}
\\\times
\Biggl[
\smallCG{L_1}{K_1}{L_2}{K_2}{L_3}{K_3}
+
\begin{closedcases}
 (-)^{L_2} \smallCG{L_1}{K_1}{L_2}{-K_2}{L_3}{K_3} & K_1\geq K_2\\
 (-)^{L_1} \smallCG{L_1}{-K_1}{L_2}{K_2}{L_3}{K_3} & K_1\leq K_2
\end{closedcases}
\Biggr]
F_{1\,K_1}(\gamma)F_{2\,K_2}(\gamma)F_{3\,K_3}(\gamma)
\Biggr]
\,\sin3\gamma\,d\gamma
.
\end{multline}
\end{widetext}

\subsection{\boldmath $\grpsochain$ Clebsch-Gordan coefficients}
\label{sec-cg}

Representations of $\grpso{5}$ couple to form new representations
according to $\grpso{5}$ Clebsch-Gordan coefficients.  If the
$\grpso{5}$ representations are labeled according to the
$\grpso{5}\<\supset\grpso{3}\<\supset\grpso{2}$ subalgebra
chain~(\ref{eqn-sochain}), each coupling coefficient may be written as
the product of an $\grpso{3}$-reduced Clebsch-Gordan coefficient and
an ordinary $\grpso{3}$ Clebsch-Gordan coefficient, according to the
Racah factorization
lemma~\cite{racah1949:complex-spectra-part4-f-shell} (see
Ref.~\cite{wybourne1974:groups} for a general discussion).  We term
the $\grpso{3}$-reduced factor an $\grpsochain$ Clebsch-Gordan
coefficient.  Coupling coefficients, such as these, which are reduced
with respect to a subalgebra are also known as ``isoscalar
factors''~\cite{edmonds1962:u3-products-isf}.

In the context of the
$\grpso{5}$ spherical harmonics, only the symmetric representations $(v,0)$ of
$\grpso{5}$ arise.  These are related by the $\grpsochain$
Clebsch-Gordan coefficients
$\smallCG{(v_1,0)}{\alpha_1 L_1}{(v_2,0)}{\alpha_2L_2}{(v_3,0)}{\alpha_3 L_3}$,
where the  $\alpha_1$, $\alpha_2$, and $\alpha_3$ are 
multiplicity indices, corresponding to the multiplicity in the $\Psi_{v\alpha{}L}$.
Let $\lbrace\chi^{(v_1,0)}_{\alpha_1L_1M_1}\rbrace$ and
$\lbrace\chi^{(v_2,0)}_{\alpha_2L_2M_2}\rbrace$ denote orthonormal bases
for $\grpso{5}$ representations of seniority $v_1$ and $v_2$,
respectively.  Then an orthonormal basis for an $\grpso{5}$
tensor-coupled product irrep of seniority $v$ is defined by
\begin{widetext}
\begin{equation}
\label{eqn-so5-cg-defn}
[\chi^{(v_2,0)}\otimes\chi^{(v_1,0)}]^{(v,0)}_{\alpha L M}
=\sum_{\substack{\alpha_1,L_1,\alpha_2,L_2\\M_1,M_2}}
\CG{(v_1,0)}{\alpha_1 L_1}{(v_2,0)}{\alpha_2L_2}{(v,0)}{\alpha
L}
\CG{L_1}{M_1}{L_2}{M_2}{L}{M}
\,
\chi^{(v_2,0)}_{\alpha_2L_2M_2}\otimes\chi^{(v_1,0)}_{\alpha_1L_1M_1}.
\end{equation}
\end{widetext}

The $\grpsochain$ Clebsch-Gordan coefficients enter into the
$\grpso{5}$ Wigner-Eckart theorem, which governs matrix elements of
$\grpsochain$ tensor operators.  The Racah factorization
lemma
may be used to write the $\grpso{5}$ Wigner-Eckart theorem in terms of
$\grpso{3}$-reduced quantities as
\begin{multline}
\label{eqn-so5-we-red}
\rme{\Psi_{v_3\alpha_3L_3}}{\hat\Psi_{v_2\alpha_2L_2}}{\Psi_{v_1\alpha_1L_1}}
\\=
\sqrt{2L_3+1}
\CG{(v_1,0)}{\alpha_1 L_1}{(v_2,0)}{\alpha_2L_2}{(v_3,0)}{\alpha_3 L_3}
\rme{\Psi_{v_3}}{\hat\Psi_{v_2}}{\Psi_{v_1}},
\end{multline}
where $\rme{\Psi_{v_3}}{\hat\Psi_{v_2}}{\Psi_{v_1}}$ is an
$\grpso{5}$-reduced (doubly-reduced) matrix element.  The factor
$\sqrt{2L_3+1}$ compensates for the corresponding factor absorbed into
the definition of the $\grpso{3}$-reduced matrix
element in~(\ref{eqn-we-defn}).   The $\grpsochain$ Clebsch-Gordan coefficients satisfy
the normalization condition
\begin{equation}
\label{eqn-cg-column}
\sum_{\substack{ \alpha_1 L_1 \\ \alpha_2 L_2}} 
\CG{(v_1,0)}{\alpha_1 L_1}{(v_2,0)}{\alpha_2 L_2}{(v_3,0)}{\alpha_3
L_3}^2 = 1
\end{equation}
from unitarity, where the phases of the Clebsch-Gordan coefficients
are real by convention.  The values for $L$ occurring within a given
$\grpso{5}$ representation $(v,0)$, and their multiplicities, are
governed by the multiplicity formula~(\ref{eqn-so5-dvL}).  The
Clebsch-Gordan coefficients vanish unless $L_1$, $L_2$, and $L_3$
satisfy the triangle inequality ($\abs{L_1-L_2}\<\leq L_3\<\leq
L_1+L_2$).  The values of $v_1$, $v_2$, and $v_3$ likewise must
satisfy the triangle inequality ($\abs{v_1-v_2}\<\leq v_3\<\leq
v_1+v_2$), as well as the parity constraint that $v_1+v_2+v_3$ be
even.

The $\grpsochain$ Clebsch-Gordan coefficient
$\smallCG{(v_1,0)}{\alpha_1 L_1}{(v_2,0)}{\alpha_2
L_2}{(v_3,0)}{\alpha_3 L_3}$ follows immediately from the
corresponding $\grpso{3}$-reduced matrix element
$\rme{\Psi_{v_3\alpha_3L_3}}{\Psi_{v_2\alpha_2L_2}}{\Psi_{v_1\alpha_1L_1}}$,
by~(\ref{eqn-so5-we-red}), once the $\grpso{5}$-reduced matrix element
$\rme{\Psi_{v_3}}{\Psi_{v_2}}{\Psi_{v_1}}$ is known.  In fact,
from~(\ref{eqn-so5-we-red}) and the normalization condition~(\ref{eqn-cg-column}),
we obtain
\begin{equation}
\label{eqn-rme-norm}
\rme{\Psi_{v_3}}{\hat\Psi_{v_2}}{\Psi_{v_1}}^2 = 
\sum_{\substack{ \alpha_1 L_1 \\ \alpha_2 L_2}} 
\frac{\rme{\Psi_{v_3\alpha_3L_3}}{\hat\Psi_{v_2\alpha_2L_2}}{\Psi_{v_1\alpha_1L_1}}^2}{2L_3+1} ,
\end{equation}
for each $L_3$ and $\alpha_3$.   This yields the
$\grpso{5}$-reduced matrix element  in terms of the summed squares of the computed $\grpso{3}$-reduced matrix elements,
to within  an arbitrary phase, which is chosen to be unity.

Explicit closed-form expressions for the $\grpso{5}$-reduced matrix
element have been given for the case
$\rme{\Psi_{v\pm1}}{\Psi_{1}}{\Psi_{v}}$, \textit{i.e.}, for the
quadrupole tensor~\cite{rowe2005:algebraic-collective}.  Moreover, by
numerical inspection of the normalization sums for an extensive set of
computed $\grpso{3}$-reduced matrix elements, we
find~\cite{rowe-INPREP} that the $\grpso{5}$-reduced matrix elements
are given by
\begin{widetext}
\begin{multline}
\label{eqn-trevor-formula}
\rme{\Psi_{v_3}}{\Psi_{v_2}}{\Psi_{v_1}}
=\frac{1}{4\pi}
\sqrt{
  \frac{
    (2v_1 + 3)(2v_2 + 3)
  }{
    (v_3 + 2)(v_3 + 1)
  }
}
\frac{
  (\tfrac12\sigma+1)!
}{
  (\tfrac12\sigma-v_1)!(\tfrac12\sigma-v_2)!(\tfrac12\sigma-v_3)!
}
\\\times
\sqrt{
  (\sigma+4)
  \frac{
    (\sigma-2v_1+1)!(\sigma-2v_2+1)!(\sigma-2v_3+1)!
  }{
     (\sigma+3)!
  }
}
,
\end{multline}
\end{widetext}
with $\sigma\<\equiv(v_1+v_2+v_3)$.  This conjectured expression has
been verified exhaustively for all combinations of values for $v_1$,
$v_2$, and $v_3$ likely to be considered in nuclear collective model
calculations.  However, we stress that is has not been proved in full
generality.  Use of the expression~(\ref{eqn-trevor-formula}) for
$\rme{\Psi_{v_3}}{\Psi_{v_2}}{\Psi_{v_1}}$ in the $\grpso{5}$
Wigner-Eckart theorem~(\ref{eqn-so5-we-red}) allows each $\grpsochain$
Clebsch-Gordan coefficients to be extracted directly from the
corresponding $\grpso{3}$-reduced matrix elements~(\ref{eqn-rme-sum}),
without the need to calculate matrix elements for all $L_1$ and $L_2$
values involved in the normalization condition~(\ref{eqn-cg-column}).
[The use of~(\ref{eqn-trevor-formula}) in the code is described in
Sec.~\ref{sec-use-instr}.]  The validity of the conjecture, for given
$v_1$, $v_2$, and $v_3$, may be established by calculating the
normalization sum for the underlying computed $\grpso{3}$-reduced
matrix elements or, equivalently, by explicitly verifying that the
extracted coefficients satisfy~(\ref{eqn-cg-column}).

The $\grpsochain$ Clebsch-Gordan coefficients obey symmetry
relations~\cite{rowe2004:spherical-harmonics}
\begin{multline}
\label{eqn-symm-12}
\CG{(v_2,0)}{\alpha_2L_2}{(v_1,0)}{\alpha_1 L_1}{(v_3,0)}{\alpha_3
L_3}
\\=
(-)^{L_1+L_2-L_3}
\CG{(v_1,0)}{\alpha_1 L_1}{(v_2,0)}{\alpha_2L_2}{(v_3,0)}{\alpha_3
L_3}
\end{multline}
and
\begin{multline}
\label{eqn-symm-13}
\CG{(v_3,0)}{\alpha_3L_3}{(v_2,0)}{\alpha_2L_2}{(v_1,0)}{\alpha_1 L_1}
=
(-)^{L_1+L_2-L_3}
\\\times
\sqrt{\frac{d_{v_1} (2L_3+1)}{d_{v_3} (2L_1+1)}}
\CG{(v_1,0)}{\alpha_1 L_1}{(v_2,0)}{\alpha_2L_2}{(v_3,0)}{\alpha_3
L_3},
\end{multline}
where $d_v=\tfrac16(v+1)(v+2)(2v+3)$ is the dimension of the
$\grpso{5}$ representation $(v,0)$, obtained from the Weyl formula 
(\textit{e.g.}, Ref.~\cite{wybourne1974:groups}).
Hence, it is only necessary to calculate
the Clebsch-Gordan coefficient for one permutation of $(v_1,v_2,v_3)$.

\section{Implementation}
\label{sec-impl}

\subsection{Overview}
\label{sec-impl-overview}

The computer code for construction of $\grpsochain$ spherical
harmonics and calculation of $\grpsochain$ Clebsch-Gordan coefficients
is implemented as a set of packages for Mathematica
6~\cite{wolfram2007:mathematica-6}.  Mathematica provides native
support for symbolic arithmetic.  In the present context, this allows
all calculations to be carried out exactly, in terms of expressions
involving square roots of rational numbers.

The basic algorithm for the calculation of $\grpsochain$
Clebsch-Gordan coefficients, as described in Sec.~\ref{sec-method},
involves three main computational tasks:
\begin{enumerate}
\item[(1)] construction of the monomials~(\ref{eqn-phi-prod}) comprising the
basis of highest-weight functions,
\item[(2)] calculation of the overlaps~(\ref{eqn-overlap-sum}) of these
monomials, as needed for the orthonormalization process, and
\item[(3)] calculation of triple overlaps~(\ref{eqn-rme-sum}), as needed for
the $\grpsochain$ Clebsch-Gordan coefficients.
\end{enumerate}
For practical implementation, two
algorithmic refinements are made to this scheme.  These
relate to the internal representation of the functions $F_K(\gamma)$
(Sec.~\ref{sec-impl-fourier}) and to the recognition (and full
utilization) of extensive redundancies among the calculations involved
in evaluating different triple overlap integrals 
(Sec.~\ref{sec-impl-me}).  Together,  an efficient treatment of these
aspects of the implementation extends the range of applicability of
the method (for calculation in exact arithmetic) from a maximal
seniority of approximately $10$ to seniorities of $\rtrim\gtrsim100$,
thereby easily yielding 
more than sufficient $\grpsochain$ Clebsch-Gordan coefficients
for nuclear structure calculations.

To understand the considerations underlying the efficient
implementation of the algorithm, let us more closely consider the
structure of the calculation.  The computationally most demanding task
is evaluation of the overlap or triple overlap integrals in the
$\Phi_{NtL}$ basis.  Recall that this involves first constructing the
function of $\gamma$ appearing in the integrand
of~(\ref{eqn-overlap-sum}) or~(\ref{eqn-rme-sum}) and then evaluating
the integral with respect to $\gamma$.  The integrand is built from
the generating functions~(\ref{eqn-phi-expansion}),
using~(\ref{eqn-phi-prod}) and~(\ref{eqn-xi-prod-stretched}).  The integrand is
thus a polynomial in the trigonometric functions $\cos\gamma$,
$\sin\gamma$, $\cos2\gamma$, $\sin2\gamma$, $\cos3\gamma$, and
$\sin3\gamma$ or, therefore, by multiple-angle identities, a
polynomial in $\cos\gamma$ and $\sin\gamma$. For the overlap
$\ol{\Phi_{N_2t_2LM}}{\Phi_{N_1t_1LM}}$, the resulting polynomial
requires powers of $\cos\gamma$ and $\sin\gamma$ as high as $N_1+N_2$,
or, for the triple overlap
$\rme{\Phi_{N_3t_3L_3}}{\hat\Phi_{N_2t_2L_2}}{\Phi_{N_1t_1L_1}}$, as
high as $N_1+N_2+N_3$.  The
integral may be evaluated exactly, but simplification of the
polynomial and integration with standard symbolic algebra software
becomes prohibitively inefficient for the construction of spherical
harmonics with $v\<\gtrsim10$.  Alternatively,
attempts at brute-force numerical integration are hampered by the
highly-oscillatory nature of such polynomials in trigonometric
functions.  Thus, it is seen that evaluation of the overlap integrals
is the defining computational challenge.

An effective solution arises from the realization that any polynomial
of degree $n$ in $\cos
\gamma$ and $\sin
\gamma$ can be decomposed as a finite sum of exponentials
\begin{equation}
\label{eqn-fourier-sum}
f(\gamma)=\sum_{k=-n}^n a_k e^{ik\gamma},
\end{equation}
\textit{i.e.},
as a finite Fourier series, of degree $n$.  Integration of exponentials
is trivial.  Such a Fourier expansion method was, in fact, used in the
final version of the code used to calculate the
Clebsch-Gordan coefficients tabulated in
Ref.~\cite{rowe2004:spherical-harmonics}. In
the present implementation, from the very beginning of the
calculation, we simply represent all $F_K(\gamma)$ coefficients,
starting with those of the integrity basis
functions~(\ref{eqn-phi-expansion}), as finite Fourier
series~(\ref{eqn-fourier-sum}).  That is, every function is
replaced by a list of Fourier coefficients $\lbrace
a_{-n},\ldots,a_{0},\ldots,a_n\rbrace$.  Then, when the integrand
in~(\ref{eqn-overlap-sum}) or~(\ref{eqn-rme-sum}) is constructed as a
function of $\gamma$, it is already manifestly Fourier expanded, and
the $\gamma$ integration is straightforward.

\subsection{\boldmath Fourier series representation of functions}
\label{sec-impl-fourier}

When the functions of $\gamma$ are represented as Fourier
sums~(\ref{eqn-fourier-sum}), there are three basic arithmetic
operations which must be carried out on the corresponding sets of
Fourier coefficients.  Addition of two functions $f(\gamma)$ and
$g(\gamma)$ is accomplished by addition of their Fourier coefficients.
Multiplication of a function by a constant is simply accomplished by
multiplication of the coefficients by that constant.  Multiplication of two functions
$f(\gamma)$ and $g(\gamma)$ by each other gives rise to a convolution
of the coefficients.  Specifically, let $f(\gamma)$ and $g(\gamma)$ be
given by Fourier sums $f(\gamma)=\sum_{r=-m}^m a_r (\eig)^r$ and
$g(\gamma)=\sum_{s=-n}^n b_s (\eig)^s$.  Expanding
the product and collecting like powers of $\eig$ yields the product
rule
\begin{equation}
\label{eqn-fourier-product-sum}
f(\gamma)g(\gamma)= \sum_{k=-(m+n)}^{m+n} c_k (\eig)^k
\end{equation}
with the new Fourier coefficients given by
\begin{equation}
\label{eqn-fourier-product-c}
c_k=
\sum_{t=\max(-2m-k,-2n+k)}^{\min(2m-k,2n+k)}a_{\frac12(k+t)}b_{\frac12(k-t)}.
\end{equation}
This is essentially the Fourier convolution theorem, in
discrete form.

For real-valued functions which are even in $\gamma$, the coefficients
$a_k$ are pure real and obey the symmetry condition $a_{-k}\<=a_{k}$.
Similarly, for odd real-valued functions, the coefficients are pure
imaginary and obey the symmetry $a_{-k}\<=-a_{k}$.  For instance, the
functions arising in the definition of the 
generating functions~(\ref{eqn-phi-expansion}) 
are $\cos m\gamma = [(\epig)^m +(\emig)^m]/2$ and 
$\sin m\gamma = [(\epig)^m - (\emig)^m]/(2i)$, which
are even and odd, respectively.  In either case, only coefficients
with $k\<\geq0$ need be stored, and only real-valued coefficients are
required, provided a factor of $i$ is absorbed into the definition of
the series for odd functions.

Thus, we use the representation
\begin{equation}
\label{eqn-fourier-symm}
f(\gamma)=
\left(\frac{1}{i}\right)^g
\sum_{k=0}^n \frac{a_k}{1+\delta_k}[(\epig)^k+(-)^g(\emig)^k],
\end{equation}
with $g\<=0$ for even functions and $g\<=1$ for odd functions.  This
is effectively a Fourier cosine series or a Fourier sine series, for
the even and odd cases, respectively.

The definite integrals needed for the calculation are of the
form $\int_0^{\pi/3} f(\gamma)\,d\gamma$.  Moreover, only odd
integrands ($g\<=1$) arise in the problem.  The definite integral of an odd
series~(\ref{eqn-fourier-symm}) on the ``sector'' $0\<\leq \gamma \<\leq
\pi/3$ is
\begin{equation}
\label{eqn-fourier-int}
\int_0^{\pi/3} f(\gamma)\,d\gamma=
\sum_{k=1}^n \frac{2a_k}{k}\Biggl[1-\cos\frac{(k\bmod 6)\pi}{3}\Biggr].
\end{equation}
Evaluation of this sum requires only a limited set of trigonometric
values, namely, $\cos k\pi/3$ ($k\<=0$, $1$, $\ldots$, $5$).

The necessary definitions for working with Fourier representations of
functions are contained in the subpackage \texttt{FourierSum}.  The
Fourier sum~(\ref{eqn-fourier-symm}) is represented symbolically by
the expression
\begin{equation*}
\text{\tt 
FourierSum[BohrGamma,$g$,\{$a_0$,$a_1$,$\ldots$,$a_n$\}] },
\end{equation*}
where $g$ indicates the symmetry ($g\<=0$ or $1$), and the $a_k$ are
the real-valued Fourier coefficients as defined
in~(\ref{eqn-fourier-symm}).  (The tag
\texttt{BohrGamma} simply serves to indicate that the
argument is
$\gamma$, since the package allows for more general possibilities.)
 Thus, for example, $\cos 2\gamma$ is represented as
\begin{equation*}
\text{\tt FourierSum[BohrGamma,0,\{0,0,1/2\}]
}.
\end{equation*}
The coefficients which arise in the present calculations are rational
numbers or square roots of rational numbers, and they are maintained
as exact symbolic expressions throughout the calculation.

The package defines addition of two \texttt{FourierSum} expressions,
multiplication by a constant, and multiplication of two
\texttt{FourierSum} expressions as extensions to the
usual Mathematica \texttt{+} and \texttt{*} operations.  The function
\texttt{IntegrateSector[$r$,$f$]} returns the integral given
in~(\ref{eqn-fourier-int}). The function
\texttt{FourierSumToTrig[$f$]} is also provided, to convert
\texttt{FourierSum} expressions back into ordinary symbolic
expressions in terms of trigonometric functions.

\subsection{Evaluation of matrix elements}
\label{sec-impl-me}

The task of computing matrix elements (triple overlap integrals) of the
$\grpsochain$ spherical harmonics is most conveniently carried out by
first calculating the matrix elements of the original monomial basis
functions $\Phi_{NtL}$, that is, the
$\rme{\Phi_{N_3t_3L_3}}{\hat\Phi_{N_2t_2L_2}}{\Phi_{N_1t_1L_1}}$.  The
desired matrix elements
$\rme{\Psi_{v_3\alpha_3L_3}}{\hat\Psi_{v_2\alpha_2L_2}}{\Psi_{v_1\alpha_1L_1}}$
in the orthonormal spherical harmonic basis then follow immediately from
the Gram-Schmidt
transformation~(\ref{eqn-gst}). It is simplest to write this using the counting
index labeling within an $L$-space, as
\begin{multline}  
\label{eqn-rme-gst}
\rme{\Psi_{L_3i_3}}{\hat\Psi_{L_2i_2}}{\Psi_{L_1i_1}} =\\
\sum_{j_1j_2j_3}T_{L_3 i_3 j_3}T_{L_2 i_2 j_2}T_{L_1 i_1 j_1}
\rme{\Phi_{L_3j_3}}{\hat\Phi_{L_2j_2}}{\Phi_{L_1j_1}}.
\end{multline}

As described in the preceding section, we have transformed the
relatively intractable integration problem into the more manageable
task of constructing products of $F_K(\gamma)$ functions in Fourier
representation.  Therefore, the computational burden now lies
primarily in evaluating the discrete Fourier
convolutions~(\ref{eqn-fourier-product-sum}--\ref{eqn-fourier-product-c}).
Although the operations involved are in principle simple arithmetic, the
number of iterations required for a single convolution is substantial,
growing as the square of the degrees of the Fourier
sums involved, and the arithmetic itself involves
symbolic simplification of expressions involving square roots of rational
numbers.

The challenge, therefore, lies in minimizing the number and
complexity of the Fourier convolutions which must be evaluated.  The
problem is best approached by noting that the matrix elements are not
to be calculated singly, but rather in aggregate, that is, as the set
of all matrix elements of an operator $\hat\Phi_{N_2t_2L_2}$ between
$L$-spaces $L_1$ and $L_3$.  The integrand in the
expression~(\ref{eqn-rme-sum}) for the matrix element is a sum over
products of $F_K(\gamma)$ functions, and products of the exact same
pairs of $F_K(\gamma)$ functions arise, redundantly, in the evaluation of many
different matrix elements.

A few straightforward observations allow us to remove these
calculational redundancies.  Most obviously, every matrix element
$\rme{\Phi_{N_3t_3L_3}}{\hat\Phi_{N_2t_2L_2}}{\Phi_{N_1t_1L_1}}$
involving the same ket $\ket{\Phi_{N_1t_1L_1}}$ and operator
$\hat\Phi_{N_2t_2L_2}$ will also involve the
same products between $F_K(\gamma)$ functions from
$\Phi_{N_1t_1L_1}(\gamma,\Omega)$ and
$\Phi_{N_2t_2L_2}(\gamma,\Omega)$.  In the notation
of~(\ref{eqn-we-overlap}), it is therefore advantageous to calculate
the coupled action of $\hat\Phi_{N_2t_2L_2}$ on any given
$\ket{\Phi_{N_1t_1L_1}}$, namely,
$[\hat\Phi_{N_2t_2L_2}\times\ket{\Phi_{N_1t_1L_1}}]^{(L_3)}$, only
once, and to reuse this intermediate result for the matrix element
with each bra $\bra{\Phi_{N_3t_3L_3}}$.

More significant, though, is the observation that all powers of
$\Phi_3\<\propto \cos3\gamma$ factor out of the summations
in~(\ref{eqn-we-overlap}).  (This will be more clearly apparent
below.)  Thus, the integrands involved in the calculation of many
different
$\rme{\Phi_{N_3t_3L_3}}{\hat\Phi_{N_2t_2L_2}}{\Phi_{N_1t_1L_1}}$,
sharing the same total label $t\<=t_1+t_2+t_3$, are actually
identical.  Even greater reduction in the number of convolutions
needed is obtained by first evaluating the integrand only for those
monomials involving no powers of $\Phi_3$
($t_1\<=t_2\<=t_3\<=0$) and only then multiplying by the relevant
power $(\cos 3\gamma)^t$.

In practice, the entire process is based on operations involving the coefficient
functions $F_K(\gamma)$ which describe functions
on $S_4$, as in~(\ref{eqn-expansion-m}).  The relevant definitions are
provided by the subpackage \texttt{WignerDSum}.
A function on $S_4$ is
represented as the expression
\begin{equation*}
\text{\tt 
WignerDSum[$L$,\{$F_0$,$F_2$,$\ldots$,$F_{\lfloor L \rfloor_2}$\}] },
\end{equation*}
where $L$ is the angular momentum and $\lfloor L \rfloor_2$ denotes
the greatest \textit{even} integer less than or equal to $L$.  The
$F_K$, in turn, are \texttt{FourierSum} expressions
(Sec.~\ref{sec-impl-fourier}).  Thus, for example, the generating
function $\Phi_1$ from~(\ref{eqn-phi-expansion}) is represented
by
\begin{equation*}
\text{\parbox{0.85\hsize}{ 
\begingroup \tt 
WignerDSum[2,\{\\
\mbox{}\quad FourierSum[BohrGamma,0,\{0,1/2\}],\\
\mbox{}\quad FourierSum[BohrGamma,1,\{0,1/2\}]
\\\}]
\endgroup .
}}
\end{equation*}
Operations of addition of two functions and multiplication by a
constant are readily defined, by entry-wise operations on the lists of
$F_K(\gamma)$ coefficients.  These operations are defined by the
package as extensions to the usual Mathematica \texttt{+} and
\texttt{*} operations.

To obtain the coupled action
$[\hat\Phi_{N_2t_2L_2}\times\ket{\Phi_{N_1t_1L_1}}]^{(L_3)}$, we
evaluate the coupled product
$[\Phi_{N_2t_2L_2}(\gamma,\Omega)\times\Phi_{N_1t_1L_1}(\gamma,\Omega)]^{(L_3)}$.
The calculation follows in a straightforward fashion from the coupling
rule~(\ref{eqn-xi-prod}) for the functions $\xi^{(L)}_K$.  The
coupling operation is provided by the \texttt{WignerDSum} package as
the function
\texttt{TensorCouple[$\Psi_1$,$\Psi_2$,$L$]}.

However, the monomials $\Phi_{N_2t_2L_2}(\gamma,\Omega)$ and
$\Phi_{N_1t_1L_1}(\gamma,\Omega)$ must first themselves be
constructed, from the definition~(\ref{eqn-phi-prod}).  As noted in
Sec.~\ref{sec-method-basis}, multiplication of highest-weight
functions is identical to stretched ($L=L_1+L_2$) tensor coupling. The
stretched product rule~(\ref{eqn-xi-prod-stretched}) is simply the
$L=L_1+L_2$ special case of the more general coupling
rule~(\ref{eqn-xi-prod}) for the $\xi^{(L)}_K$.  Therefore, the
$\Phi_{NtL}$ are also constructed in the code by use of this same function
\texttt{TensorCouple}, as
\begin{equation}
\label{eqn-phi-tensor}
\Phi_{NtL}=
\Bigl[
\bigl[\Phi_1^{(2)}\bigr]^{n_1}
\<\times
\bigl[\Phi_2^{(2)}\bigr]^{n_2}
\<\times
\bigl[\Phi_3^{(0)}\bigr]^{n_3}
\<\times
\bigl[\Phi_4^{(3)}\bigr]^{n_4}
\Bigr]^{(L)},
\end{equation}
where $\Phi_1^{(2)}$ is the tensor with highest-weight component
$\Phi_1(\gamma,\Omega)$, \textit{etc.}, and $L\<=2n_1+2n_2+3n_4$ (Sec.~\ref{sec-method-basis}).
To complete the calculation of the matrix element, the overlap
with $\bra{\Phi_{N_3t_3L_3}}$ must be evaluated. In fact, the function
\texttt{TensorCouple} also suffices for this purpose.  The sum of
products of $F_K(\gamma)$ arising in the overlap
integral~(\ref{eqn-overlap-sum}) is readily expressed in coupled form.
By the coupling rule~(\ref{eqn-xi-prod}), 
\begin{equation}
[\Psi^{(L)}_2 \times \Psi^{(L)}_1]^{(0)}_0
=\frac{2}{(2L+1)^{1/2}} \sum_{\substack{K=0\\\text{even}}}^L
F_{1\,K}(\gamma)F_{2\,K}(\gamma).
\end{equation}
This follows from the basic expression for $L\<=0$ Clebsch-Gordan
coefficients, much like the usual scalar product result $[T^{(L)}_2
\times T^{(L)}_1 ]^{(0)}_0 \<= (-)^L (2L+1)^{-1/2}\sum_M (-)^M
T^{(L)}_{1\,M}T^{(L)}_{2\,-M}$.

It remains then to carry out the
integration over $\gamma$, by use of \texttt{IntegrateSector} (Sec.~\ref{sec-impl-fourier}).
If we define
$\langle f(\gamma)\rangle  \equiv 8\pi^2 \int_0^{\pi/3} f(\gamma)\,\sin 3\gamma \,
d\gamma$,
where the factor of $8\pi^2$ results from the integration over Euler
angles, then the reduced matrix element deduced from the
overlap~(\ref{eqn-we-overlap}) is simply
\begin{multline}
\label{eqn-rme-tensor}
\rme{\Phi_{N_3t_3L_3}}{\Phi_{N_2t_2L_2}}{\Phi_{N_1t_1L_1}}  
\\
=
\langle [\Phi_{N_3t_3L_3}\times(\Phi_{N_2t_2L_2}\times\Phi_{N_1t_1L_1})^{(L_3)}]^{(0)}_0 \rangle.
\end{multline}

Expressing the overlap in terms of a coupled product does not change
the underlying calculation (and at most affords a convenient economy
of coding).  However, this formulation does permit the factorization
of $(\cos3\gamma)^t$ from the integrand to be expressed in an
especially symmetric (and explicit) form.
Observe that $\Phi^{(0)}_3 \propto \cos3\gamma$ is a scalar, and, therefore, unlike the other
generating functions, it factors out of the coupled product
defining the monomial in~(\ref{eqn-phi-tensor}), and subsequently out
of the couplings in the zero-coupled product
in~(\ref{eqn-rme-tensor}). Hence, if we label each of the monomials by
its exponents with respect to each of the generating functions, the
integral factorizes as
\begin{multline}
\label{eqn-tau-extract}
\bigl[ 
\Phi_{[n'_1,n'_2,n'_3,n'_4]}^{(L_3)}
\times
\bigl(
\Phi_{[\nu_1,\nu_2,\nu_3,\nu_4]}^{(L_2)}
\times
\Phi_{[n_1,n_2,n_3,n_4]}^{(L_1)}
\bigr)^{(L_3)}
\bigr]^{(0)}_0
\\
=
\bigl[ 
\Phi_{[n'_1,n'_2,0,n'_4]}^{(L_3)}
\times
\bigl(
\Phi_{[\nu_1,\nu_2,0,\nu_4]}^{(L_2)}
\times
\Phi_{[n_1,n_2,0,n_4]}^{(L_1)}
\bigr)^{(L_3)}
\bigr]^{(0)}_0
\\\times
\Phi_{[0,0,t,0]},
\end{multline}
where $t\<=n_3'+\nu_3+n_3$ is the \textit{total} exponent of
$\Phi_3^{(0)}$ occurring in the product.  Note that only monomials of
the form $\Phi_{[n_1,n_2,0,n_4]}$, with $n_3\<=0$, are actually needed
in the calculation, along with
$\Phi_{[0,0,t,0]}\<=(\sqrt{2}\cos3\gamma)^t$.

The mechanics of the optimized calculation of matrix elements
therefore proceed as follows.  The monomials $\Phi_{[n_1,n_2,0,n_4]}$
are constructed by~(\ref{eqn-phi-tensor}).  To eliminate redundancy,
this is done recursively, that is, by multiplication of the monomial
of one lower degree in $n_1$, $n_2$, or $n_4$ by the appropriate
generating function, and all the intermediate results are cached.
Then the successive coupled products in~(\ref{eqn-tau-extract}) are constructed and cached:
$\bigl(
\Phi_{[\nu_1,\nu_2,0,\nu_4]}^{(L_2)}
\times
\Phi_{[n_1,n_2,0,n_4]}^{(L_1)}
\bigr)^{(L_3)}
$ (stage I),
$\bigl[ 
\Phi_{[n'_1,n'_2,0,n'_4]}^{(L_3)}
\times
\bigl(
\Phi_{[\nu_1,\nu_2,0,\nu_4]}^{(L_2)}
\times
\Phi_{[n_1,n_2,0,n_4]}^{(L_1)}
\bigr)^{(L_3)}
\bigr]^{(0)}_0
$ (stage II), and finally the full integrand with $\Phi_{[0,0,t,0]}$
included (stage III).
\begin{figure}
\begin{center}
\includegraphics*[width=\hsize]{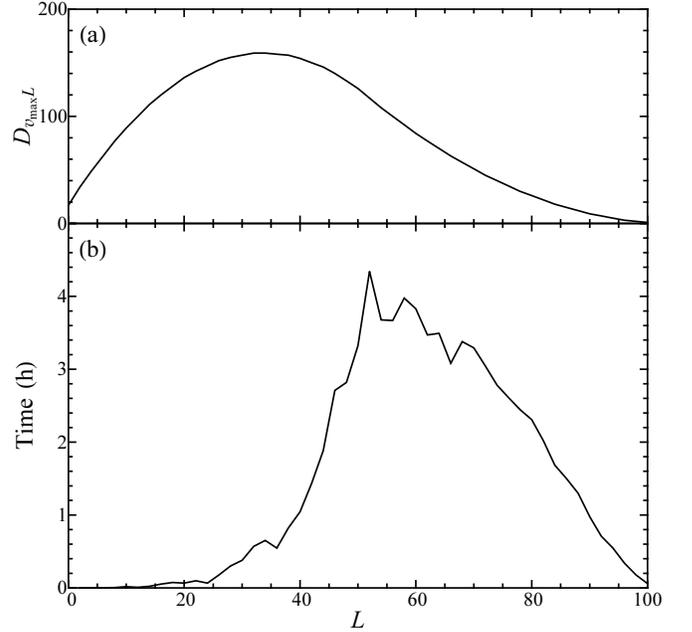}
\end{center}
\caption{Basis sizes and corresponding execution times, for calculating the matrix
elements of $\Psi_{112}\<\propto\calQ$ between the spherical harmonics
with $v\<\leq50$.  (a)~The number of basis functions $D_{\vmax L}$
($\vmax\<=50$), for $L$ even.  (b)~Time for evaluating the matrix
elements within a single $L$-space ($L_1\<=L_3\<\equiv L$), again only
shown for
$L$ even.  (The basis dimensions and, consequently, execution times
are significantly smaller for $L$ odd.)  All times are for
calculations carried out under  Mathematica~6 for Linux, running on a
$2.2\,\mathrm{GHz}$ Advanced Micro Devices Opteron processor.}
\label{fig-profile}
\end{figure}

For matrix elements within an $L$-space, Hermiticity reduces the
number of independent matrix elements nearly by half.  Since, under
complex conjugation, $\Phi_{NtL}$ satisfies $\Phi_{NtLM}^*=
(-)^{L-M}\Phi_{NtL-M}$, it follows~\cite{edmonds1960:am} that
\begin{multline}
\label{eqn-rme-adjoint}
\rme{\Phi_{N_3t_3L_3}}{\hat\Phi_{N_2t_2L_2}}{\Phi_{N_1t_1L_1}} 
\\
= (-)^{L_3+L_2-L_1}
\rme{\Phi_{N_1t_1L_1}}{\hat\Phi_{N_2t_2L_2}}{\Phi_{N_3t_3L_3}}^*.
\end{multline}

As a specific example of the computational savings provided by the
caching of intermediate products, consider the calculation of matrix
elements of $\Phi_{112}\<\propto\calQ$ within the space
$L_1\<=L_3\<=40$, for all monomials of degree $\rtrim\leq50$. [This
calculation yields the $\grpsochain$ Clebsch-Gordan coefficients with
$v_2\<=1$ and $L_2\<=2$, for all $v_1$ and $v_3\<\leq50$.]  There are
$D_{50,40}\<=154$ basis functions and thus $23\,716$ matrix elements
to be calculated, or $11\,935$ after the
relation~(\ref{eqn-rme-adjoint}) is taken into account.  Evaluation of
the coupled action of $\Phi_{112}$ (stage I) involves only $21$
couplings, since the coupling is only carried out for basis functions
with $n_3\<=0$.  Evaluation of the scalar couplings (stage~II)
constitutes the bulk of the calculation.  Whereas each of the couplings
evaluated in stage~I involves a low-$L$, low-degree factor
($\Phi_{112}$), in stage~II both factors,
$\Phi_{[n'_1,n'_2,0,n'_4]}^{(L_3)}$ and $\bigl(
\Phi_{[\nu_1,\nu_2,0,\nu_4]}^{(L_2)}
\times
\Phi_{[n_1,n_2,0,n_4]}^{(L_1)}
\bigr)^{(L_3)}
$, are typically of high $L$ (hence each contains many $F_K$ terms) and
high degree with respect to $\gamma$ (hence many terms are involved in the
Fourier sums).  A total of $231$ such couplings are necessary.
 Many
more distinct expressions, namely, $3001$, must be evaluated at stage
III, but each entails only a single multiplication of scalar functions
(Fourier convolution) and is therefore not
computationally intensive.  

Total execution times for evaluation of
these matrix elements within each $L$-space, as well as the dimensions
of these spaces, are shown in Fig.~\ref{fig-profile}.  It is seen that
calculations for higher $L$ are essentially more demanding than those
for lower $L$, even if the bases for the $L$-spaces are of the same
size.  For higher $L$, a larger number of $F_K$ terms are involved in
each calculation.  Furthermore, the basis monomials tend to involve higher
powers of those generating functions which carry angular momentum
($\Phi^{(2)}_1$, $\Phi^{(2)}_2$, and $\Phi^{(3)}_4$), in preference to
$\Phi^{(0)}_3$, lessening the extent to which the optimizations
discussed above can simplify the calculation.

The full, optimized process of evaluating the matrix
elements of a monomial $\Phi_{N_2t_2L_2}$ between two
$L$-spaces $L_1$ and $L_3$ is carried out by the function
\texttt{CalculateMonomialMatrix}.  The considerations just
described also apply to eliminating redundancies in the evaluation of
the overlaps needed for the orthonormalization process.  

\section{Use of the computer code}
\label{sec-use}

\subsection{Installation}
\label{sec-use-install}

The Computer Physics Communications Program Library deposit associated
with this article contains Mathematica package files
\texttt{Gamma\textrm{\-}Harmonic.m},
\texttt{Fourier\textrm{\-}Sum.m}, \texttt{WignerD\textrm{\-}Sum.m},
\texttt{Gram\textrm{\-}Schmidt.m}, and \texttt{Cache.m},  in ASCII text format.  
Annotated code is included
for the packages in both Mathematica notebook format and
Portable Document Format.  The deposit also includes a Mathematica
notebook demonstrating the use of the code (\texttt{Example.nb}) and
example output files containing tabulations of Clebsch-Gordan
coefficients (see Sec.~\ref{sec-use-instr}).

All package files must be placed in a directory in the Mathematica
file search path, as explained in the Mathematica
documentation~\cite{wolfram2007:mathematica-6}, or else in the current
working directory for Mathematica.  The package
\texttt{GammaHarmonic} must then be loaded, by evaluating
\texttt{Get["GammaHarmonic`"]}, as demonstrated in
\texttt{Example.nb}.

\subsection{Instructions}
\label{sec-use-instr}
\DeclareRobustCommand{\fcntableentry}[2]{\tt #1
&
\begin{minipage}[t]{0.5\hsize}
\raggedright #2
\end{minipage}
}
\begin{table*}
\caption{Control functions for $\grpsochain$ spherical harmonic and
$\grpsochain$ Clebsch-Gordan coefficient generation.}
\label{tab-control}
\begin{ruledtabular}
\renewcommand{\arraystretch}{1}
\begin{tabular}{l@{~}l}
Function  & Description\\
\hline
\fcntableentry{ConstructQNSet[$\vmax$,$\Lmax$]}{
Constructs a list of monomial $(N,L,t,K)$ labels for each $L$-space.
}\\
\fcntableentry{ConstructMonomials[$\vmax$,$\Lmax$]}{
Precaches the $\Phi_{NtL}$ constituting the monomial basis.}\\
\fcntableentry{ConstructMonomialMatrices[$\vmax$,$\Lmax$,\{$L_2$,$i_2$\}]}{ 
Calculates the matrix elements 
$\rme{\Phi_{L_3i_3}}{\hat\Phi_{L_2i_2}}{\Phi_{L_1i_1}}$ in the monomial basis,
by evaluation of the triple overlap integrals.
}\\
\fcntableentry{ConstructBasisGST[$\vmax$,$\Lmax$] }{
Constructs the 
Gram-Schmidt transformation
coefficients.
}\\
\fcntableentry{ConstructBasisMatrices[$\vmax$,$\Lmax$,\{$L_2$,$i_2$\}] }{ 
Calculates the matrix elements 
$\rme{\Psi_{L_3i_3}}{\Psi_{L_2i_2}}{\Psi_{L_1i_1}}$ in the spherical
harmonic basis, by Gram-Schmidt transformation of the monomial matrix elements.
}\\
\fcntableentry{WriteSO5CGTable[$\vmax$,$\Lmax$,\{$L_2$,$i_2$\},{\rm\it
filename}] }{ 
Writes a tabulation of $\grpsochain$ Clebsch-Gordan coefficients,
as defined in the text, including both the exact rational and
floating-point values.  
}\\
\fcntableentry{ReadSO5CGTable[$\vmax$,$\Lmax$,\{$L_2$,$i_2$\},{\rm\it
filename}] }{ 
Reads back a tabulation of $\grpsochain$ Clebsch-Gordan coefficients,
as defined in the text.  
}
\end{tabular}
\end{ruledtabular}
\end{table*}


The \texttt{GammaHarmonic} package provides calculation control
functions to carry out the
several steps involved in constructing and tabulating a set of
$\grpsochain$ Clebsch-Gordan coefficients.
The syntax for each of the control functions is given in Table~\ref{tab-control}.
Here, we provide basic instructions for carrying out these calculational
tasks.  A
worked example, with sample input and output, is given in the
Mathematica notebook
\texttt{Example.nb}.

(1) The labeling scheme for the basis monomials must first be set up, using
\texttt{ConstructQNSet} (see
Table~\ref{tab-control} for the appropriate syntax).  This function
assembles and stores a list of $(N,L,t,K)$ labels for the monomials
($N\<\leq\vmax$) spanning each $L$-space
($L\<\leq\Lmax$)~\cite{fn-gst-order}. 

(2) The necessary monomials $\Phi_{NtL}$ for the calculation
must be constructed, with 
\texttt{Construct{\rmh}Monomials}.  Only monomials with $n_3\<=0$ 
are evaluated, for the reasons discussed in Sec.~\ref{sec-impl-me}.

(3) The function
\texttt{Construct{\rmh}Monomial{\rmh}Matrices} is then used to construct
matrix elements $\rme{\Phi_{L_3i_3}}{\hat{\Phi}_{L_2i_2}}{\Phi_{L_1i_1}}$ in
the monomial basis, by the process discussed in
Sec.~\ref{sec-impl-me}.  The function calculates all matrix elements
for a given $L_2$ and $i_2$, between all $L$-spaces  $L_1$ and
$L_3$, up to $\Lmax$, subject to the triangle inequality.  This
calculation should be carried out for any $(L_2,i_2)$ which
contribute, in~(\ref{eqn-rme-gst}), to the final spherical harmonic
matrix elements of interest.  For nuclear structure applications,
these would typically include $(L_2,i_2)\<=(2,1)$, for the electric quadrupole
operator, and $(L_2,i_2)\<=(0,2)$, needed for $\cos^n3\gamma$
contributions to the potential in the Hamiltonian.  The calculation
must also be carried out for $(L_2,i_2)\<=(0,1)$, \textit{i.e.}, for
the identity operator $\Phi_{010}$, since this provides the overlaps
needed for the Gram-Schmidt orthonormalization.

(4) The function \texttt{ConstructBasisGST} must be used to
generate the Gram-Schmidt transformation coefficients $T_{Lij}$.  

(5) Finally, the function
\texttt{ConstructBasisMatrices} is used to calculate 
the reduced matrix elements
$\rme{\Psi_{L_3i_3}}{\hat{\Psi}_{L_2i_2}}{\Psi_{L_1i_1}}$ of the spherical
harmonics, by Gram-Schmidt transformation~(\ref{eqn-rme-gst}) of
the previously-calculated monomial matrix elements.  

The $\grpsochain$ Clebsch-Gordan coefficients follow from these
computed $\grpso{3}$-reduced matrix elements, by the $\grpso{5}$
Wigner-Eckart theorem~(\ref{eqn-so5-we-red}) and the normalization
condition~(\ref{eqn-cg-column}), as described in Sec.~\ref{sec-cg}.
Individual Clebsch-Gordan coefficients may be accessed 
with the function
\texttt{SO5ClebschGordan[$\vmax$,\linebreak[0]\{$v_1$,$L_1$,$\alpha_1$\},\linebreak[0]\{$v_2$,$L_2$,$\alpha_2$\},\linebreak[0]\{$v_3$,$L_3$,$\alpha_3$\}]}.
So that the coefficients can be extracted independently, without
requiring computation of matrix elements for all $L_1$ and $L_2$
values encountered in the normalization sum~(\ref{eqn-cg-column}), the
function \texttt{SO5ClebschGordan} directly applies the 
expression~(\ref{eqn-trevor-formula}) for the $\grpso{5}$-reduced
matrix element in the $\grpso{5}$
Wigner-Eckart theorem.

The function
\texttt{WriteSO5CGTable} outputs a tabulation of $\grpsochain$ Clebsch-Gordan coefficients
sharing the same $v_2$, $\alpha_2$, and $L_2$.  These correspond to
spherical harmonic matrix elements sharing the same
$\hat{\Psi}_{v_2\alpha_2L_2}$.
Coefficients are listed in order of increasing $L_3$, $L_1$, $v_3$,
$\alpha_3$, $v_1$, and $\alpha_1$ (that is, with the last of these
indices varying most rapidly).  Only coefficients with $L_1\<\leq L_3$
are tabulated, in recognition of the symmetry
relation~(\ref{eqn-symm-13}), and only those coefficients for which
nonzero values are allowed under the $\grpso{3}$ triangle inequality and $\grpso{5}$
selection rules (Sec.~\ref{sec-cg}) are included.

The Clebsch-Gordan coefficients are written both exactly and as
floating point numbers.  The exact value of a Clebsch-Gordan
coefficient can be expressed as the signed square root of a rational
number, $\pm\sqrt{a/b}$.  The magnitude of each value is squared in
the output, to eliminate the need for radicals, so the value written
is $\pm a/b$.  Because the numerators and denominators appearing in
these rational numbers can be large (exceeding 400 decimal digits, for
instance, for $\vmax\<=50$), the floating point value is also given.
This is meant to facilitate input by external programs written in
languages which do not provide native support for arbitrary-length
integers.  Each row of the tabulation has the form
\begin{equation*}
v_1~L_1~\alpha_1~v_2~L_2~\alpha_2~v_3~L_3~\alpha_3
~~x~~\pm a\mathtt{/}b,
\end{equation*}
where $x$ is the floating point representation of $\pm a/b$.

Tabulated coefficients may be read back 
with \texttt{ReadSO5CGTable}.  The corresponding $\grpso{3}$-reduced matrix
elements are recovered and stored as matrices
\texttt{Basis{\rmh}Operator{\rmh}Matrix[$\vmax$,\linebreak[0]\{$L_2$,$i_2$\},\linebreak[0]\{$L_3$,$L_1$\}]},
which may then be used, for instance, for ACM calculations in
Mathematica.
When using the matrix elements calculated by the code, it should be
noted that the factor of $8\pi^2$ which arises from the integration
over Euler angles has been suppressed in the
actual calculations.  (This choice is made in order to eliminate the
symbol $\pi$ from input and output.)  If the
true normalization is desired, all computed matrix elements must be
multiplied by $(8\pi^2)^{-1/2}$.

Two example tabulations of $\grpsochain$ Clebsch-Gordan coefficients
are included with the Computer Physics Communications Program Library
deposit.  These contain coefficients with $(v_2,L_2)\<=(1,2)$ and 
$(3,0)$ [that is, $(L_2,i_2)\<=(2,1)$ and $(0,2)$],
for all seniorities $v_1$ and $v_3$ up to $\vmax\<=50$ and for all allowed
values of $L_1$ and $L_3$ at these seniorities ($\Lmax\<=100$).  The
tabulations are given in the files
\texttt{basis-50-100-cg-2-1.dat} and \texttt{basis-50-100-cg-0-2.dat}.
(The numerical labels indicate $\vmax$, $\Lmax$, $L_2$, and
$i_2$, respectively.)  The
tabulated coefficients are of sufficient extent to support basic
nuclear structure calculations with the ACM.

Intermediate results may be saved to files at various stages of the
calculation, and later retrieved, allowing computations to be resumed
or extended~\cite{fn-resumption} at a later time.  In particular, monomial matrix elements
may be saved with
\texttt{Write{\rmh}Monomial{\rmh}Matrices} and subsequently read back with
\texttt{Read{\rmh}Monomial{\rmh}Matrices} (see \texttt{Example.nb} and
the internal program
documentation).     
Similar functions are defined to write and read back the Gram-Schmidt
transformation coefficients and to write out the $\grpso{3}$-reduced
matrix elements of the spherical harmonics in matrix format.

\subsection{\boldmath Explicit expressions for the $\grpsochain$ spherical harmonics}
\label{sec-use-explicit}

Explicit expressions for the $\grpsochain$ spherical harmonics, as
functions of $\gamma$ and the Euler angles (Table~\ref{tab-psi}), are
not needed for diagonalization of the collective model Hamiltonian or
calculation of  transition matrix elements in the ACM framework.  The
$\grpsochain$ Clebsch-Gordan coefficients suffice.  However, if ACM
calculations are carried out in an $\grpsutimes$ basis, probability
distributions $P(\beta,\gamma)$ for the wave functions in coordinate
space can then be obtained by combining the known radial wave
functions $R^\lambda_n(\beta)$~\cite{rowe2005:algebraic-collective}
and the expressions for the $\Psi_{v\alpha{}L}(\gamma,\Omega)$.  The
general procedure is given in the appendix of
Ref.~\cite{caprio2005:axialsep},  and example probability distributions
may be found in Figs.~3 and~4 of that reference.

The function
\texttt{ConstructBasisWaveFunctions\hideargs{[$\vmax$,$\Lmax$]}}
explicitly constructs the spherical harmonics
$\Psi_{v\alpha{}L}(\gamma,\Omega)$, by taking linear combinations of
the monomials $\Phi_{NtL}$, according to the Gram-Schmidt
transformation~(\ref{eqn-gst}).  The spherical harmonics may be
converted from Fourier representation to symbolic expressions
involving trigonometric functions by use of
\texttt{FourierSumToTrig} (Sec.~\ref{sec-impl-fourier}), as demonstrated in
\texttt{Example.nb}.  A normalization factor of $(8\pi^2)^{-1/2}$ must be restored in these
results, as indicated in Table~\ref{tab-psi}, if
the true normalization is desired.

\subsection{Algebraic collective model infrastructure}
\label{sec-use-acm}

If ACM calculations are
to be carried out within Mathematica, the
\texttt{GammaHarmonic} package provides several additional functions which
can facilitate this process.  As already noted
(Sec.~\ref{sec-use-instr}), the package provides the necessary
function (\texttt{ReadSO5CGTable}) for input of previously-calculated
$\grpsochain$ Clebsch-Gordan coefficients.
The function 
\texttt{TruncateBasisMatrices} may then be used to truncate the stored
matrices of $\grpso{3}$-reduced matrix elements to lower $\vmax$, in
order to reduce the product space dimensions for ACM calculations.
The
\texttt{GammaHarmonic} package also defines several functions (\texttt{SO5{\rmh}LSpace{\rmh}Size}, \texttt{SO5{\rmh}Branching},
\texttt{SO5{\rmh}LSpace{\rmh}Seniorities}, \textit{etc.}) based on the
$\grpsochain$ dimension and branching formulas of
Appendix~\ref{app-so5-dim} (see the internal program documentation).
These functions are of use in indexing the $\grpsutimes$ basis
functions and constructing the seniority contribution to the kinetic
energy operator.

\section{Conclusion}
\label{sec-concl}

The computational methods described in this article, as implemented in
the accompanying computer code, make possible the calculation of a
sufficient set of $\grpsochain$ Clebsch-Gordan coefficients to support
fully converged nuclear structure calculations with the algebraic
collective model (ACM).  The $\grpsutimes$ algebraic structure of the
ACM basis permits matrix elements of an essentially unlimited set of
potential and kinetic energy operators to easily be constructed.  The
Bohr collective model in the resulting calculational scheme is thus
genuinely an \textit{algebraic} collective model.  The availability of
$\grpsochain$ Clebsch-Gordan coefficients, in conjunction with a large
body of analytic expressions for $\beta$ matrix
elements~\cite{rowe2005:algebraic-collective,rowe2005:radial-me-su11},
makes it possible to algebraically construct the matrix elements for
any collective model Hamiltonian expressible as a polynomial in the
collective quadrupole moments and canonical momenta.  Algebraic
expressions for matrix elements of a variety of
other operators, including the term $\beta^{-2}$ occurring in the
Davidson
potential~\cite{elliott1986:gsoft-davidson,rowe1998:davidson}, have
also been derived~\cite{rowe2005:algebraic-collective,rowe2005:radial-me-su11}.

The calculated $\grpsochain$ Clebsch-Gordan coefficients now permit the
diagonalization of the Bohr Hamiltonian for potentials of essentially
arbitrary $\gamma$ stiffness.  This allows application of the ACM to the full
range of nuclear quadrupole rotational-vibrational structure, from
spherical oscillator to axial rotor to triaxial rotor.  With an
appropriate optimized choice of $\beta$ basis
functions~\cite{rowe2005:algebraic-collective}, fully converged
calculations can be carried out very efficiently, providing a valuable
tool for studying collective motion in nuclei.

The $\grpsutimes$ framework also
opens the door for more detailed exploration of the formal relationship
between the Bohr collective model and the interacting boson model
(IBM)~\cite{iachello1987:ibm}, through the
$\grpu{6}\<\supset\grpu{5}\<\supset\grpso{5}$ and
$\grpu{6}\<\supset\grpso{6}\<\supset\grpso{5}$ bases.  Algebraic
collective model methods have already been
applied~\cite{rowe2005:ibm-geometric} to calculation of collective
states within the $\grpso{6}$ limit of the IBM.  The availability
$\grpsochain$ Clebsch-Gordan coefficients enables such studies to be
enhanced and extended.

\begin{acknowledgments}
We thank J.~Repka for helpful discussion.  This work was supported by
the US DOE under grant DE-FG02-95ER-40934 and by the Natural Sciences
and Engineering Research Council of Canada.
\end{acknowledgments}

\appendix

\section{\boldmath Branching and multiplicity for the reduction $\grpsochain$}
\label{app-so5-dim}

The multiplicity of each $\grpso{3}$ irrep $(L)$ within the
$\grpso{5}$ irrep $(v,0)$ is needed in the code, both to determine the
$\grpsochain$ branching for the basis and to convert between the $(v\alpha{}L)$ 
and $(Li)$ labeling schemes.  The result of
Refs.~\cite{kishimoto1971:boson-cfp,eisenberg1987:v1}  can be
expressed more compactly as
\begin{equation}
\label{eqn-so5-dvL}
d_{vL}=
(\lfloor\tfrac13(v - b)\rfloor+1)\theta_{v-b}
-\lfloor \tfrac13(v - L + 2)\rfloor\theta_{v-L+2},
\end{equation}
where $b\<\equiv L/2$ for $L$ even or $(L+3)/2$ for $L$ odd.  The step
function $\theta_k$ is unity for $k\<\geq0$ and zero otherwise.
The
 dimension $D_{\vmax L}\<\equiv\sum_{v=0}^\vmax d_{v L}$ 
of
the basis for a given $L$-space, truncated at seniority $v\<\leq\vmax$, is consequently
\begin{multline}
\label{eqn-so5-DvL}
D_{\vmax L}=\bigl[ f(\vmax-b,3) + (\vmax-b+1)\bigr]\theta_{\vmax-b}
\\
-f(\vmax-L+2,3)\theta_{\vmax-L+2},
\end{multline}
where $f(n,m)\<\equiv\sum_{k=0}^n\lfloor k/m \rfloor$ is given by
\begin{equation}
f(n,m)=\lfloor n/m \rfloor \bigl[ n+1 - \tfrac12 m(\lfloor n/m
\rfloor+1) \bigr].
\end{equation}

\vfil


\providecommand{\ELSEVIER}{}
\ELSEVIER\newcommand{\identity}[1]{{#1}}



\begin{thebibliography}{44}
\bibitem{fn-degree}
\bibinfo{note}{Strictly speaking, the degree of an arbitrary polynomial in the
  components $\calQ_m$ of the quadrupole tensor is not well-defined. Since
  $\calQ \cdot \calQ \<\equiv \sum_m \abs{\calQ_m}^2\<= 1$, a polynomial may be
  multiplied by arbitrary factors of $\calQ \cdot \calQ$ with no effect. The
  definition of degree must therefore be restricted to polynomials that cannot
  be reduced to polynomials of lower degree by means of the identity $\calQ
  \cdot \calQ\<=1$. This condition is automatically satisfied by the
  highest-weight functions obtained as products~(\ref{eqn-phi-prod}) of
  generating functions.}
\bibitem{fn-gst-order-foreshadow}
\bibinfo{note}{See, however, endnote~\cite{fn-gst-order} regarding the ordering
  of monomials of the same $L$ and $N$.}
\bibitem{fn-resumption}
\bibinfo{note}{If calculations have already been completed through
  $\Lmax\<=\Lmaxnaught$, they may be continued from this point onward (to
  higher $\Lmax$) by specifying the option \texttt{Last->$\Lmaxnaught$} to any
  of the calculation control functions (Table~\ref{tab-control}) or
  input/output functions. Alternatively, by this mechanism, calculations for
  different sets of $L$-spaces may be carried out concurrently on multiple
  processors or systems.}
\end{thebibliography}
\end{document}